\documentclass[aps,twocolumn,prb,superscriptaddress,floatfix,showpacs,citeautoscript,longbibliography]{revtex4-2}

\usepackage{amsmath}
\usepackage{amssymb}
\usepackage{graphicx}
\usepackage{epsf}
\usepackage{epsfig}
\usepackage{braket}
\usepackage{bm}
\usepackage{amsfonts}
\usepackage{color,soul}
\usepackage{times}
\usepackage[utf8]{inputenc}

\usepackage{float}
\usepackage{verbatim}
\usepackage{mathtools}
\usepackage[dvipsnames,svgnames,table]{xcolor}
\usepackage{hyperref}
\usepackage{babel}
\usepackage{orcidlink}
\usepackage{enumerate}
\hypersetup{
pdfstartview={FitH},
colorlinks=true,    
linkcolor=NavyBlue, 
citecolor=Blue,   
filecolor=NavyBlue, 
urlcolor=NavyBlue   
}

\newcommand{\ve}{\varepsilon}

\newcommand{\al}{\alpha}
\newcommand{\bt}{\beta}
\newcommand{\sg}{\sigma}
\newcommand{\up}{\uparrow}
\newcommand{\dn}{\downarrow}

\newcommand{\la}{\langle}
\newcommand{\ra}{\rangle}

\newcommand{\rl}{\frac{\pi}{a_{x}}}
\newcommand{\mt}[1]{\mathrm{#1}}
\newcommand{\mc}[1]{\mathcal{#1}}

\begin{document}
\title{Electronic and magnetic properties of graphene-fluorographene nanoribbons: Controllable semiconductor-metal transition}
\author{R.~M.~{Guzm\'an-Arellano}\orcidlink{0000-0003-4210-7511}}
\affiliation{Facultad de Ciencias F\'{\i}sicas, Universidad Nacional Mayor de San Marcos, 15081 Lima, Peru}
\author{A.~D.~{Hern\'andez-Nieves}\orcidlink{0000-0001-5345-4522}}
\affiliation{Centro At{\'{o}}mico Bariloche and Instituto Balseiro,
Comisi\'on Nacional de Energ\'{\i}a At\'omica (CNEA)- Universidad Nacional de Cuyo (UNCUYO), 8400 Bariloche, Argentina}
\affiliation{Consejo Nacional de Investigaciones Cient\'{\i}ficas y T\'ecnicas (CONICET), Argentina}
\author{F.~M.~Peeters\orcidlink{0000-0003-3507-8951}}
\affiliation{Nanjing University of Information Science and Technology, Nanjing 210044, China}
\affiliation{Department of Physics, University of Antwerp, Groenenborgerlaan 171, B-2020 Antwerpen, Belgium}
\author{Gonzalo Usaj\orcidlink{0000-0002-3044-5778}}
\affiliation{Centro At{\'{o}}mico Bariloche and Instituto Balseiro,
Comisi\'on Nacional de Energ\'{\i}a At\'omica (CNEA)- Universidad Nacional de Cuyo (UNCUYO), 8400 Bariloche, Argentina}
\affiliation{Instituto de Nanociencia y Nanotecnolog\'{i}a (INN-Bariloche), CNEA-CONICET, Argentina}
\affiliation{CENOLI, Universit\'e Libre de Bruxelles - CP 231, Campus Plaine, B-1050 Brussels, Belgium}

\begin{abstract}
We investigate the electronic and magnetic properties of graphene channels ($2$--$4$~nm wide) embedded within fluorographene, focusing on two distinct interfaces: the fully fluorinated $\al$ interface and the half-fluorinated $\bt$ interface. Density functional theory (DFT) calculations reveal that $\al\al$ systems exhibit semiconducting behavior with antiferromagnetic ordering, closely resembling pristine zigzag graphene nanoribbons. In contrast, $\al\bt$ systems display ferromagnetism and a width-dependent semiconductor-to-metal transition. To enable the study of larger systems, we develop and validate effective Hubbard models for both $\al\al$ and $\al\bt$ channels. Building upon DFT results and a Wannier function analysis, these models accurately reproduce the electronic structure and magnetic ordering observed in DFT calculations.  Crucially, our $\al\bt$ model successfully captures the semiconductor-to-metal transition.   Application of this model to larger systems reveals the persistence of a ferromagnetic state with spin polarization localized at the $\al$ edge. Our results demonstrate the potential of fluorination for targeted property engineering and provide a basis for exploring graphene-fluorographene systems in device applications ranging from microelectronics to spintronics.
\end{abstract}
\maketitle

\section{Introduction}
Graphene nanoribbons (GNRs) hold a significant promise for semiconductor applications due to their tunable width-dependent energy band gaps \cite{kusakabe:2003,Son2006a, kan:2008,Son2006, han:2010,jiao:2010,Tian:2023,Wang:2021}. Recent experimental advances have enabled the fabrication of atomically precise GNRs \cite{rizzo:2020,Yamaguchi:2020,Way:2022}, opening new avenues for device exploration \cite{Jiang:2023}.
Alternatives approaches for creating graphene nanostructures involve selective hydrogenation or fluorination of graphene sheets, or carving GNRs within graphane or fluorographene \cite{sessi:2009, balog:2010, withers:2011}, strategies that have  been successfully demonstrated experimentally. For example, reversible local modification of graphene's electronic properties was achieved using controlled hydrogen passivation with a scanning tunneling microscope tip \cite{sessi:2009}. This technique enabled the formation of nanoscale graphene patterns. Furthermore, patterned absorption of atomic hydrogen onto specific sites within graphene's Moiré superlattice has been shown to induce a band gap, as confirmed by both experiments and \textit{ab initio} calculations \cite{balog:2010}.

Fluorination techniques also offer compelling possibilities. Electron beam irradiation can selectively transform insulating fluorinated graphene into conducting or semiconducting graphene \cite{withers:2011}. Additionally, thermo-chemical nanolithography allows for the fabrication of chemically isolated GNRs as narrow as $40$~nm \cite{lee:2013}. Most recently, a reversible electron beam activation technique was used to directly ``write'' semiconducting/insulating superlattices of fluorographene channels with high resolution ($9$-$15$~nm) \cite{Li:2020}.

These experimental developments have spurred substantial theoretical interest in hybrid nanostructures \cite{singh:2009,li:2009,hernandez:2010,lu:2009}. Studies have demonstrated that the energy band gaps of zigzag graphane nanoribbons increase with decreasing width \cite{li:2009,lu:2009}. \textit{Ab initio} calculations have also revealed that the band gap of free-standing hybrid graphene/graphane and graphene/fluorographene nanoribbons is primarily determined by the graphene region \cite{hernandez:2010,lu:2009,tang:2011}.

In this work, we investigate the interplay of interfacial fluorination, electronic properties, and magnetic behavior in zigzag graphene channels carved on fluorographene. Density functional theory (DFT) calculations and a complementary Anderson-Hubbard (AH) mean-field model are employed. We analyze two different fluorination levels at the graphene/fluorographene interfaces: (i) a fully fluorinated zigzag chain ($\al$ interface) and (ii) a half-fluorinated zigzag chain ($\bt$ interface). Our results reveal that the interface composition strongly influences the electronic and magnetic properties of the graphene/fluorographene superlattice, even in the undoped (neutral) state.

The paper is organized as follows: Section II describes our DFT methodology. Sections III and IV present the crystalline and electronic structures of the zigzag graphene channels. In Sec. V, we use Wannier90 calculations to derive a tight-binding Hamiltonian, capturing the essential electronic states near the Fermi level. Section VI builds an AH model from these insights, reproducing key graphene nanoribbon properties. Section VII validates the AH model against DFT results for wider graphene channels, allowing us to confidently extend our studies to system sizes that are computationally prohibitive for DFT calculations.  Finally, Sec. VIII provides a summary and future perspectives.

\begin{figure}[t]
\includegraphics[width=0.9\columnwidth]{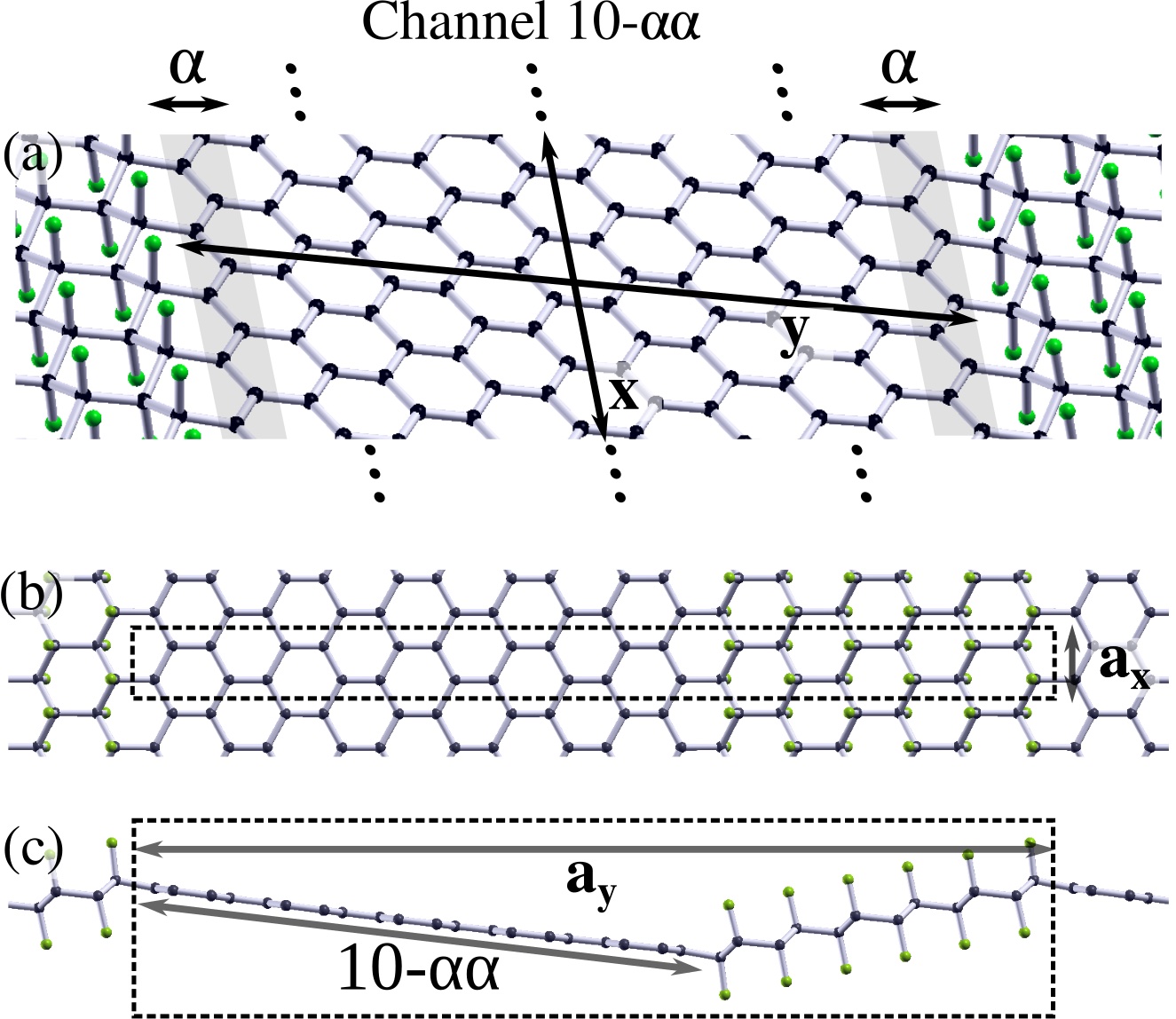}
\caption{\label{relaxed_aa} Relaxed crystalline structures of the graphene nanoribbons with fluorographene interfaces. These systems are graphene channels carved on fluorographene. The graphene-fluorographene $\al$ interfaces are enclosed by dashed lines in (a). (b), (c) The $10$-$\al\al$ unit cell along the (b) $x$ longitudinal  and (c) $y$ transversal directions. Green balls are F atoms.}
\end{figure}
\section{Computational approach}
We performed density functional theory (DFT) calculations using the Quantum Espresso (QE) package \cite{QE} to investigate the electronic and magnetic properties of graphene channels. The Perdew-Burke-Ernzerhof (PBE) exchange-correlation functional \cite{pbe} was employed, along with a plane-wave basis set for electronic wave functions and charge density. Energy cutoffs of $70$ and $420$~Ry were applied, respectively. We used ultrasoft pseudopotentials \cite{Vanderbilt1990} to describe ion-electron interactions.
Brillouin zone (BZ) sampling was tailored to the system's electronic structure: (i) semiconductors, uniform $24\times 1\times 1$ $k$-point mesh; (ii) metals, uniform $36\times 1\times 1$ $k$-point mesh; (iii) near semiconductor-metal transition, finer uniform $54\times 1\times 1$ $k$-point mesh. Gaussian smearing was applied, with a degauss value of $0.005$~Ry for systems away from the transition and $0.001$~Ry near the transition. Crystalline structures were relaxed until forces and stress on atoms were below $0.001$~eV/\AA~ and $0.5$~GPa, respectively.
\section{Crystalline structure of the graphene channels}
We study two types of zigzag graphene channels carved on fluorographene, denoted as $n$-$\al\al$ and $n$-$\al\bt$. Here, $n$ indicates the number of zigzag chains in the graphene channel, while $\al$ and $\bt$ represent distinct graphene-fluorographene interfaces \cite{hernandez:2010,Schmidt2010}.
\begin{figure}[t]
\includegraphics[width=0.9\columnwidth]{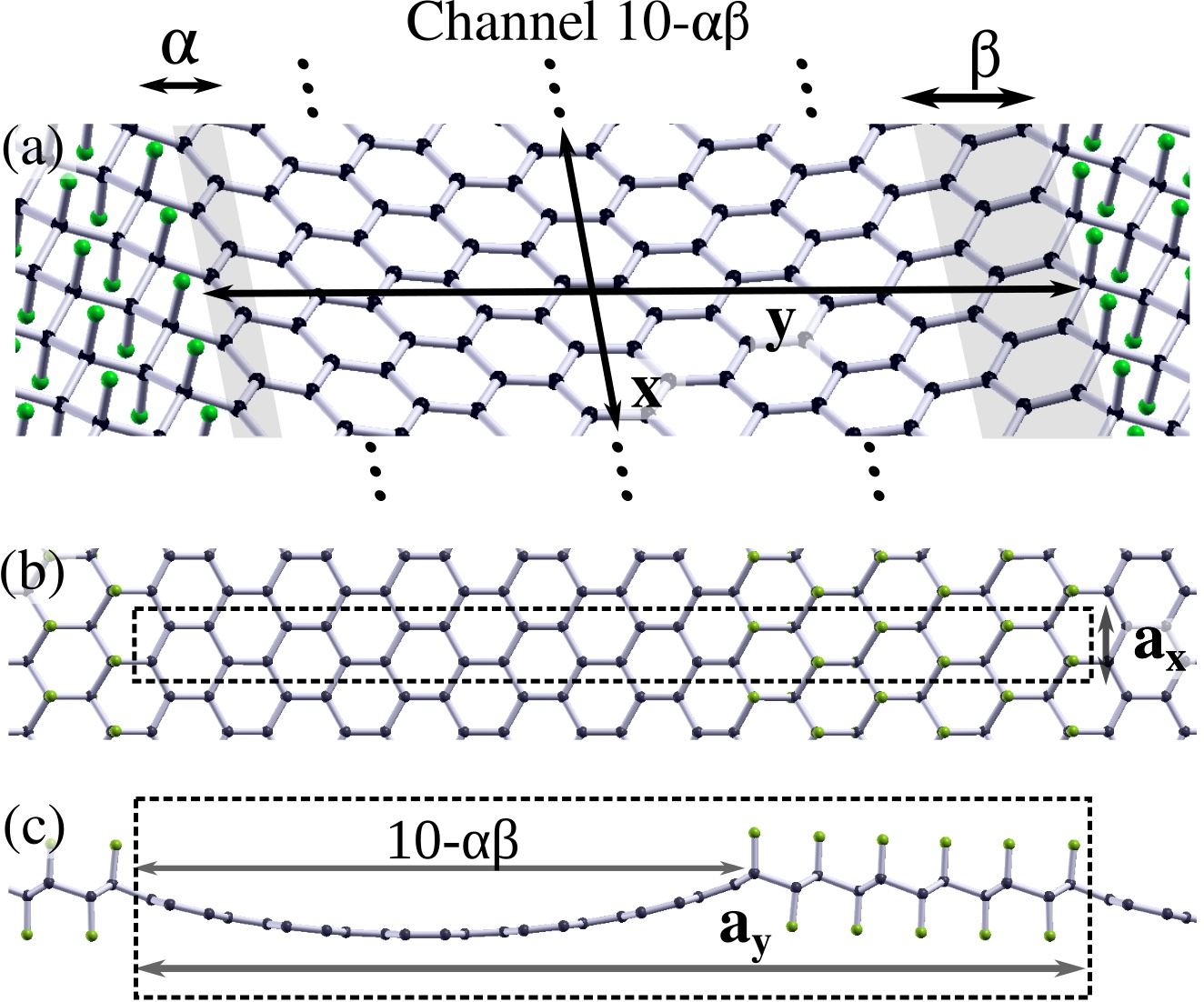}
\caption{\label{relaxed_ab} 
Relaxed crystalline structures of $n$-$\al\bt$ graphene-fluorographene nanoribbons ($n=10$), similar to Fig.~\ref{relaxed_aa} but with $\al\bt$ interfaces. The graphene channel interfaces are labeled as $\al$ (fully fluorinated, left) and $\bt$ (half-fluorinated, right). The dashed lines in (a) highlight these contrasting interfaces.}
\end{figure}
%
Figures \ref{relaxed_aa} and \ref{relaxed_ab} illustrate these interfaces (highlighted by shaded regions). The $\al$  interface exhibits no fluorination, all carbon atoms of the bordering graphene zigzag chain are not bonded to fluorine atoms. In contrast, the $\bt$ interface displays partial fluorination, with alternating carbon atoms along the bordering zigzag chain bonded to fluorine atoms. Figures \ref{relaxed_aa}(b) and \ref{relaxed_aa}(c), as well as Figs. \ref{relaxed_ab}(b) and \ref{relaxed_ab}(c), show top and side views of the structures. The unit cell repetition, due to the periodic boundary conditions, generates an array of alternating graphene channels, which are separated by fluorographene regions. 

The crystalline structures reveal differing hybridizations within the systems. Carbon atoms in the graphene channels retain sp$^2$  hybridization, while fluorographene regions exhibit sp$^3$ hybridization. This difference drives the contrasting structural features at the interfaces. In the n-$\al\al$ channels, interfacial bonds create an alternating up-down distortion: the bonds along the interfaces push up on one edge and push down on the other one, resulting in a planar graphene region. The n-$\al\bt$ channels, on the other hand, experience downward forces at both edges due to bonding patterns, leading to a curved graphene channel. The fluorographene in both cases remains planar, demonstrating its higher rigidity compared to graphene.

Fluorographene's rigidity also influences the longitudinal lattice parameter ($\mt{a_{x}}$), which remains approximately $\mt{a_{x}}=2.52$~\AA~across all studied systems. This value aligns with experimental findings \cite{nair:2010}. The transversal lattice parameter ($\mt{a_{y}}$), however, varies based on the graphene channel width. All systems feature fluorographene regions with six fluorinated zigzag chains, providing sufficient separation to isolate the electronic states of adjacent channels near the Fermi level.  Tables \ref{table_aa} and \ref{table_ab} detail the specific $\mt{a_{y}}$ values for each system.
\begin{table}[b]
\begin{center}
\begin{tabular}{|l|c|c|c|c|}
\hline 
Channel & $\mt{a_{y}}$ [\AA] & $M_{abs}$ [$\mu_{B}$] & $M_{\al}$ [$\mu_{B}$] & $\Delta$ [eV] \\ \hline
 $6$-$\al \al$ & $25.85$ & $1.10$ & $0.30$ & $0.47$ \\ \hline
 $8$-$\al \al$ & $30.09$ & $1.16$ & $0.30$ & $0.38$ \\ \hline
$10$-$\al \al$ & $34.33$ & $1.23$ & $0.30$ & $0.31$ \\ \hline
\hline
\end{tabular}
\caption{Properties of the $n$-$\al\al$ nanoribbons obtained from DFT calculations. The transversal lattice parameter ($\mt{a_{y}}$), the absolute magnetization ($M_{abs}$), and the energy gap ($\Delta$) depend on channel width ($n$), while the edge magnetization ($\mt{M_{\al}}$) stays fixed.}
\label{table_aa}
\end{center}
\end{table}
\section{Electronic structure of graphene channels}
\begin{figure}[t]
\includegraphics[width=0.95\columnwidth]{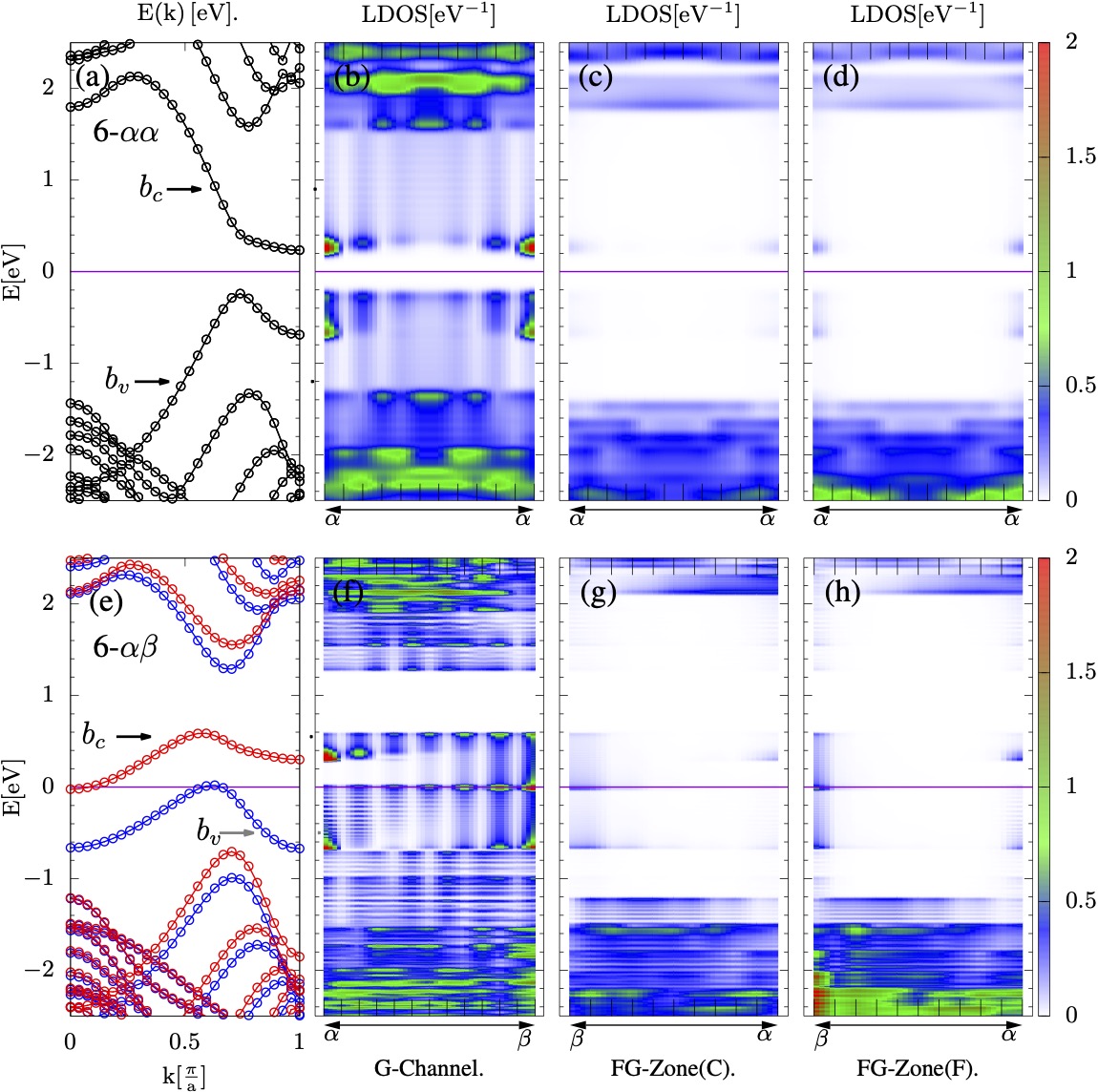}
\caption{ 
Band diagram ($\mt{E_{k}}$) and local density of states (LDOS) for (a)-(d) $6$-$\al\al$ and (e)-(h) $6$-$\al\bt$  graphene nanoribbons. The arrowheads in the LDOS plots indicate the positions of the $\al$ and $\bt$ interfaces. The $6$-$\al\al$ channel exhibits a semiconducting band gap ($\Delta=0.47$~eV), while the $6$-$\al\bt$ channel is near a semiconductor-metal transition.  Near the Fermi level, the LDOS is mainly concentrated in the graphene (G) regions and has negligible contribution in the middle of the fluorographene (FG) regions, highlighting the insulating nature of fluorographene and its role in suppressing transverse electronic excitations, and leading to $k_{y}$-independent band structures in this energy range.}
\label{LDOS_6}
\end{figure}

Figure \ref{LDOS_6} reveals distinct electronic properties for the  $6$-$\al\al$ and $6$-$\al\bt$ graphene-fluorographene superlattices. A key finding is that the type of interface between the graphene channel and fluorographene regions dramatically impacts the electronic properties. The $6$-$\al\al$ exhibits a clear semiconducting gap, while the $6$-$\al\bt$ system is near a semiconductor-metal transition. This striking difference demonstrates that the type of interfacial fluorination provides a powerful mechanism to control the electronic properties of these superlattices.
\begin{table}[b]
\begin{center}
\begin{tabular}{|l|c|c|c|c|c|}
\hline 
Channel & $\mt{a_{y}}$ [\AA] & $M_{abs}$ [$\mu_{B}$] & $M_{t}$ [$\mu_{B}$]& $M_{\bt}$ [$\mu_{B}$]&  $M_{\al}$ [$\mu_{B}$]  \\ \hline
 $4$-$\al \bt$ & $21.75$ & $1.60$ & $0.97$ & $0.38$ & $0.34$ \\ \hline
 $6$-$\al \bt$ & $25.83$ & $1.47$ & $0.84$ & $0.31$ & $0.31$ \\ \hline
 $8$-$\al \bt$ & $30.06$ & $0.82$ & $0.45$ & $0.08$ & $0.31$ \\ \hline
$10$-$\al \bt$ & $34.29$ & $0.71$ & $0.37$ & $0.03$ & $0.31$ \\ \hline 
\hline
\end{tabular}
\caption{Properties of $n$-$\al\bt$ graphene-fluorographene nanoribbons from DFT calculations.  The transversal lattice parameter ($\mt{a_{y}}$) increases with channel width ($n$), while the longitudinal lattice parameter ($\mt{a_{x}}$) remains constant at $2.52$~\AA~due to fluorographene's rigidity.  The total magnetization ($M_{t}$) and the absolute magnetization ($M_{abs}$) decrease with increasing channel width. Notably, the $\bt$ edge magnetization ($M_{\bt}$) diminishes towards zero for wider channels, while the $\al$ edge magnetization ($M_{\al}$) remains constant at $0.31$~$\mu_{B}$.}
\label{table_ab}
\end{center}
\end{table}

The local density of states (LDOS) analysis of Figs.~\ref{LDOS_6}(b)-\ref{LDOS_6}(d) and Figs.~\ref{LDOS_6}(f)-\ref{LDOS_6}(h) confirms that these states are primarily localized within the graphene channel, with negligible contributions in the central region of the fluorographene. This supports our treatment of the channels as electronically isolated within the fluorographene and validates the observed $k_{y}$-independence of the band structure near the Fermi level. This isolation arises from fluorographene's large bandgap, with states residing well above and below the Fermi level. Therefore, for the purpose of our study, these systems can be accurately modeled as individual graphene channels embedded in bulk fluorographene.

These findings underscore the remarkable control achievable through interfacial fluorination. By selectively modifying the $\al$ and $\bt$ interfaces, we can precisely tailor the electronic properties of the graphene channel between semiconducting and metallic behavior. This opens up potential avenues for designing graphene-based devices with tunable electronic properties.
\subsection{ $\al\al$ channels}
\begin{figure}[t]
\includegraphics[width=0.95\columnwidth]{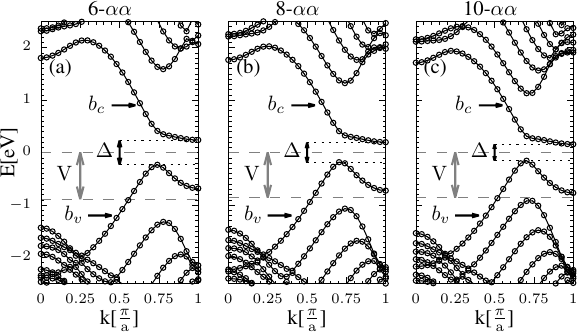}
\caption{Spin-polarized band structure of $n$-$\al\al$ gra\-phe\-ne-fluorographene systems for $n=6, 8,$ and $10$. Conduction and valence bands are labeled as $b_{v}$ and $b_{c}$, respectively. These $n$-$\al\al$ nanoribbons exhibit semiconducting behavior, with the band gap ($\Delta$) decreasing as the number of zigzag chains ($n$) in the graphene channel increases. In this case, bands of opposite spin states coincide.}
\label{bands_AA} 
\end{figure}

Figure \ref{bands_AA} presents the spin-polarized band structures of $n$-$\al\al$ graphene-fluorographene superlattices for $n=6, 8$, and $10$. The presence of an indirect energy gap ($\Delta$) between the valence ($b_{v}$) and conduction ($b_{c}$) bands confirms their semiconducting nature. Crucially, the gap magnitude decreases with channel width ($n$), ranging from approximately $0.47$~eV for $n=6$ to $0.31$~eV for $n=10$; see Table \ref{table_aa}. This width-dependent band gap highlights the potential of these systems for tunable electronic properties, relevant for various semiconductor applications \cite{kusakabe:2003,Son2006a, kan:2008,Son2006, han:2010,jiao:2010,Tian:2023,Wang:2021}.

\begin{figure}[t]
\includegraphics[width=0.9\columnwidth]{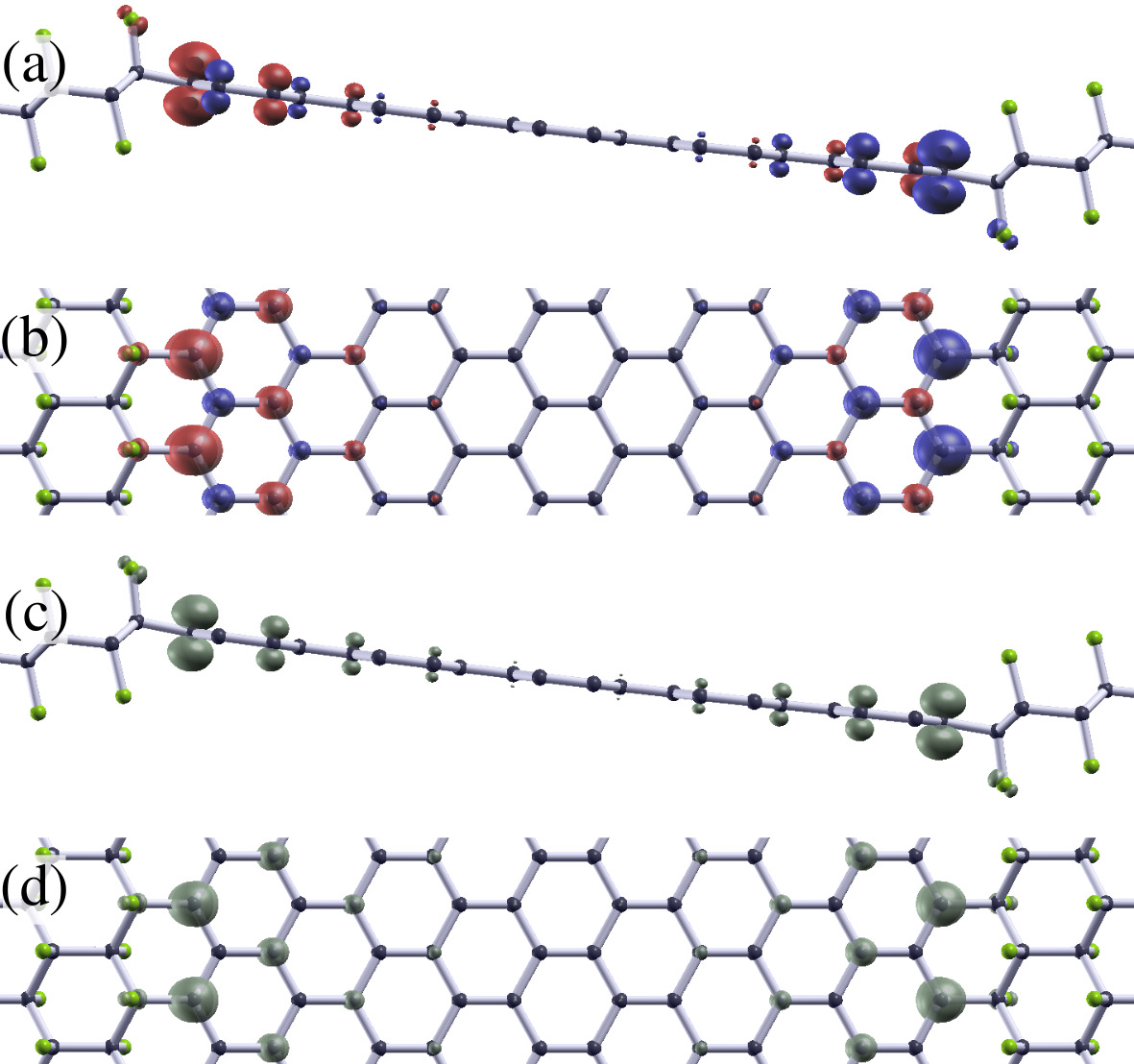}
\caption{\label{SP_AA} Spin density polarization, $m(\bm{r})=\rho_{\up}(\bm{r}) - \rho_{\dn}(\bm{r})$, of the $10$-$\al\al$ graphene-fluorographene system. (a) Side view: red and blue lobules represent opposite spin polarizations. (b) Top view: notice the antiferromagnetic order along the transverse direction to the interfaces and ferromagnetic order along the longitudinal direction. The system is globally antiferromagnetic (AF) with higher spin polarization at the channel edges. (c), (d) Integration of the local density of states (ILDOS) within the energy window ($\mt{V}$) defined in Fig.~\ref{bands_AA}(c). These states, located below the Fermi energy, exhibit $\mt{p_{z}}$ character, with higher weight at the channel edges.}
\end{figure}

Figures \ref{SP_AA}(a) and \ref{SP_AA}(b) reveal the spin density polarization, $m(\bm{r})=\rho_{\up}(\bm{r}) - \rho_{\dn}(\bm{r})$, within the graphene channel of the $10$-$\al \al$ system.  This polarization exhibits a globally antiferromagnetic (AF) order: antiferromagnetic along the transverse direction to the interface and ferromagnetic along the longitudinal direction. The system's net magnetization ($M_t$) is zero, resulting in spin-degenerate bands. This ordering is clearly visualized in the top view [Fig.~\ref{SP_AA}(b)].

To understand the nature of the occupied states near the Fermi level, we calculated the integrated local density of states (ILDOS) within the energy window $\mt{V}$ marked in Fig.~\ref{bands_AA}(c). This analysis, shown in Figs.~\ref{SP_AA}(c) and \ref{SP_AA}(d), reveals a dominant $\mt{p_{z}}$ orbital character, with higher weight at the channel edges.  This indicates that the observed spin polarization originates from these $\mt{p_{z}}$states, with spin-up ($\up$) and spin-down ($\dn$) states localized on different sublattices. Specifically, sublattice A (B) at the left (right) edge is preferentially occupied by $\up$ ($\dn$) states.

Table \ref{table_aa} summarizes key electronic, magnetic, and structural properties of $n$-$\al\al$ systems. The fluorographene regions enforce a fixed longitudinal lattice parameter $\mt{a_{x}}=2.52$~\AA, while the transversal parameter ($\mt{a_{y}}$) increases with channel width. Despite the presence of spin polarization, all systems exhibit zero global magnetization ($M_{t}=0$) due to the antiferromagnetic order. The absolute magnetization $M_{abs}$ [the integrated absolute value of $m(\bm{r})$] increases with $n$, while the energy gap ($\Delta$) decreases. While the edge states spread more as the channel becomes wider, in agreement with the reduction of the energy gap, the magnetization at the edges, $M_{\al}=\pm 0.30$~$\mu_{B}$, remains the same.
\subsection{$\al\bt$ channels}
\begin{figure}[b]
\includegraphics[width=0.95\columnwidth]{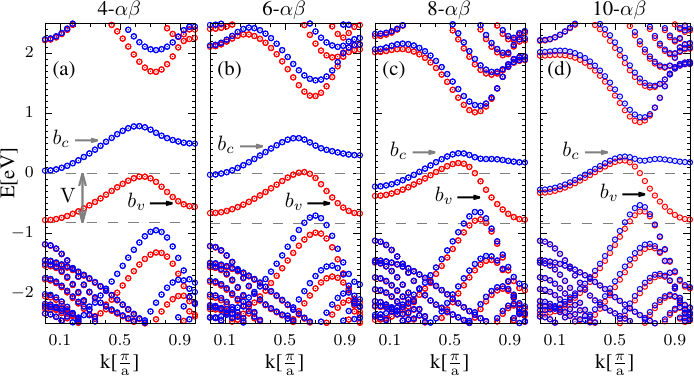}
\vspace{0cm} \caption{\label{bands_AB} Spin-polarized band structure of $n$-$\al\bt$ graphene-fluorographene systems for $n=4,6,8$ and $10$. Red and blue dots represent opposite spin states ($\uparrow$ and $\downarrow$). The ferromagnetic ground state leads to a spin-split band structure around the Fermi level. The band structure also reveals a semiconductor-metal transition with increasing channel width. At $\mt{k}=0$, the energy gap between the conduction ($b_{c}$) and valence ($b_{v}$) bands decreases with increasing $n$.}
\end{figure}
\begin{figure}[t]
\includegraphics[width=0.9\columnwidth]{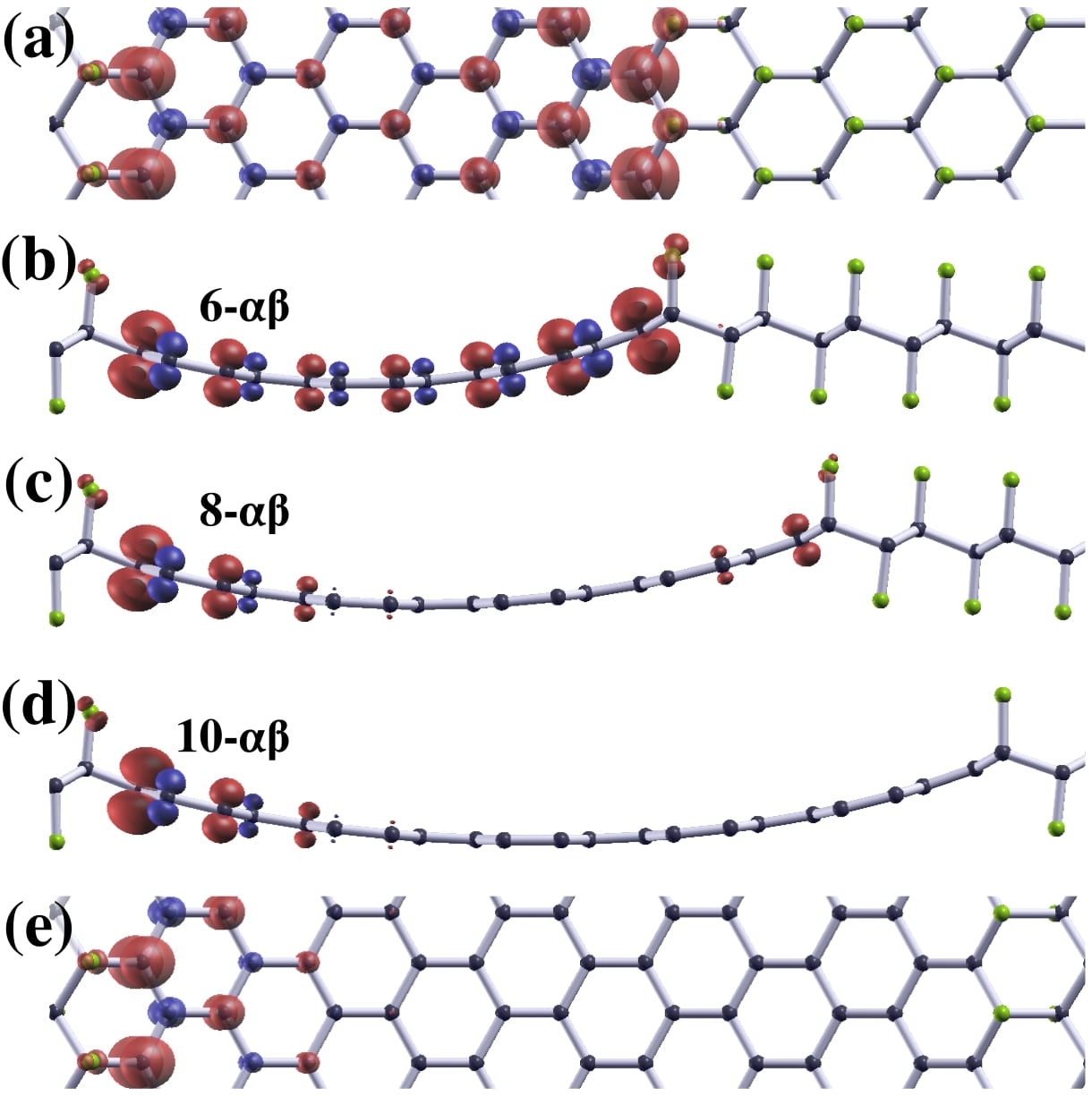}
\vspace{0cm} \caption{\label{SP_AB} Three-dimensional plot of the spin density polarization, $m(\bm{r})$, of the $n$-$\al\bt$ systems for $n = 6, 8$, and $10$.  (a), (e) Top views; (b)--(d) side views. The spin polarization is higher at the channel edges and decreases towards the center, with red and blue lobules representing opposite spin polarizations. The weight at the $\al$ edge remains constant, while it weakens at the $\bt$ edge with increasing channel width ($n \geq 8$) and nearly vanishes for (d), (e) $n = 10$. Top views (a) and (e) reveal a ferromagnetic order along the longitudinal direction and an antiferromagnetic order along the transverse direction. Despite this, the net magnetization is nonzero, making the $n$-$\al\bt$ nanoribbons ferromagnetic.}
\end{figure}
\begin{figure}[t]
\includegraphics[width=0.9\columnwidth]{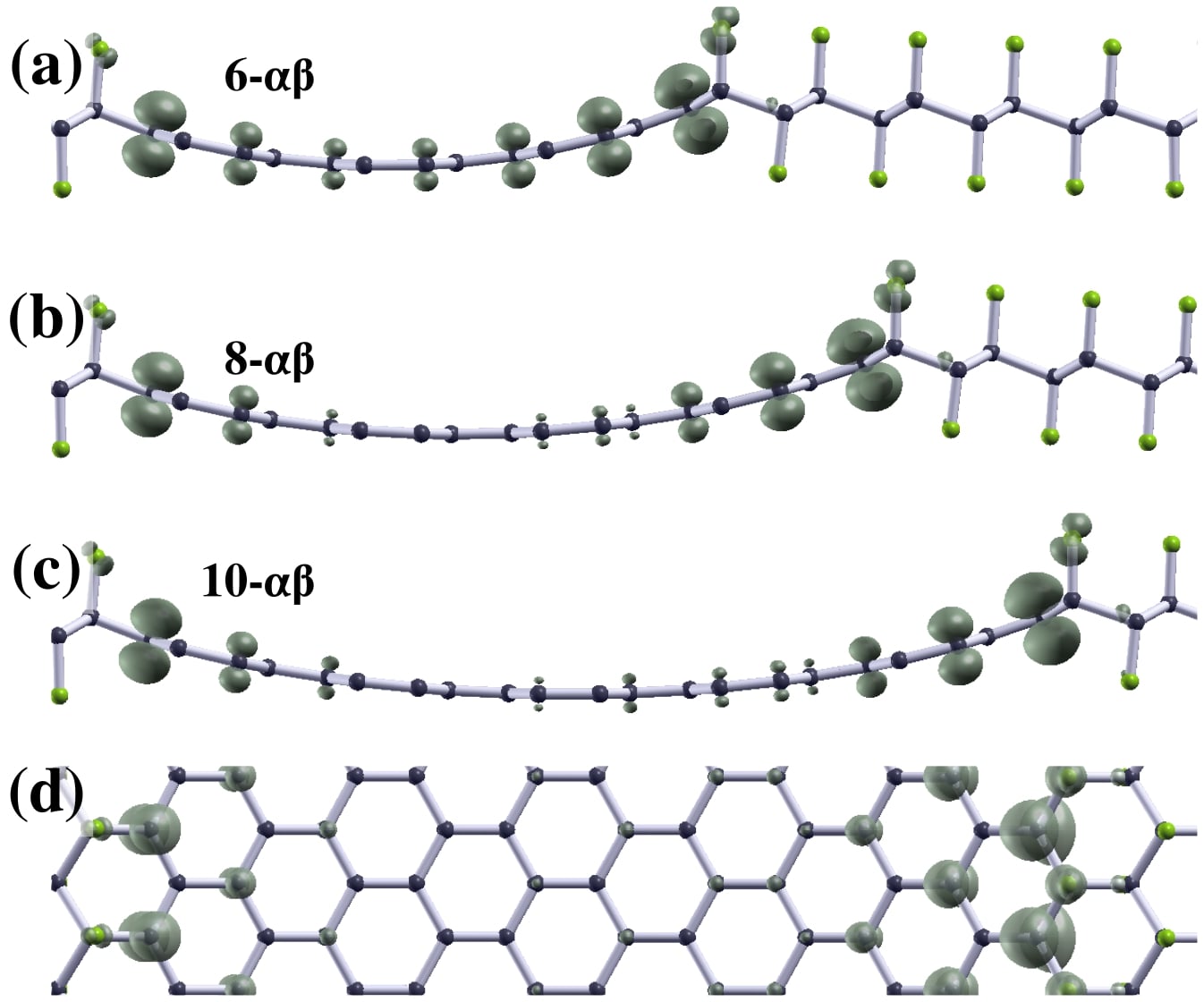}
\caption{\label{EE_AB} Integrated local density of states (ILDOS) within the energy window ($\mt{V}$) defined in Fig. \ref{bands_AB} for $\al\bt$ graphene-fluorographene systems. Similar to the $\al\al$ systems [Figs. \ref{SP_AA}(c) and \ref{SP_AA}(d)], the $\al\bt$ states near the Fermi level exhibit $\mt{p_{z}}$ character, with their weight concentrated along the graphene channel.}
\end{figure}
The spin-polarized electronic structure of $n$-$\al\bt$ nanoribbons, with $n = 4, 6, 8$ and $10$, are shown in Fig.~\ref{bands_AB}. The spin $\up$ ($\dn$) band is plotted with red (blue) color. The distinct spin splitting highlights the ferromagnetic ground state. As $n$ increases, this splitting decreases at $\mt{k}\sim0$, signaling a transition from semiconducting ($n<$6) to metallic ($n>$6) behavior, while the band splitting does not change significantly for $\mt{k}\sim\mt{\rl}$. This semiconductor to metal transition is linked to the occupation of the conduction band ($b_{c}$) near $\mt{k}\sim0$ and the consequent partial occupation of the valence band ($b_{v}$).

A comparison of the band structures of $n$-$\al\al$ (Fig.~\ref{bands_AA}) and $n$-$\al\bt$ (Fig.~\ref{bands_AB}) reveals that the most significant differences, without considering the spin splitting, occur at $\mt{k}\sim0$. This indicates that the states of the valence and conduction bands near $\mt{k}\sim0$ are strongly influenced by the distinct chemical nature of the $\bt$ interface,  the key structural difference between the two systems.

Figure~\ref{SP_AB} illustrates the spin density polarization, $m(\bm{r})$, of $n$-$\al\bt$  graphene-fluorographene superlattices. Spin polarization is concentrated within the graphene channels and exhibits unbalanced antiferromagnetic ordering in the transverse direction. This results in a global ferromagnetic (FM) state with finite net magnetization. Interestingly, spin polarization at the $\bt$ interface weakens with increasing channel width, nearly vanishing for $n=10$ [Figs.~\ref{SP_AB}(d) and \ref{SP_AB}(e)]. This behavior correlates with the spin-minority occupation of the $b_{c}$ band near $\mt{k}\sim0$, as seen in Fig.~\ref{bands_AB}. Furthermore, Fig.~\ref{bands_AB} reveals that the states below the Fermi level near $\mt{k}\sim0$ are nearly spin-degenerated for $n\geq10$, confirming that these states have minimal contributions in the spin-polarized regions.

Figure \ref{EE_AB} shows the ILDOS within the energy window $\mt{V}$ marked in Fig.~\ref{bands_AB}. For the $6$-$\al\bt$ nanoribbon [Fig.~\ref{EE_AB}(a)], the ILDOS strongly resembles the spatial profile of the spin density polarization [Fig.~\ref{SP_AB}(b)], indicating that the states within the window $\mt{V}$ are the primary  contributors to the spin-polarized regions. For the wider $8$-$\al\bt$ and $10$-$\al\bt$ systems, however, a notable difference arises at the $\bt$ edge between the ILDOS [Figs.~\ref{EE_AB}(b) and \ref{EE_AB}(c)] and the spin density polarization [Figs.~\ref{SP_AB}(c) and \ref{SP_AB}(d)] due to the reduced spin polarization of that edge, in agreement with the evolution of the band structure near $\mt{k} \sim 0$ (Fig.~\ref{bands_AB}) with increasing channel width.
Table \ref{table_ab} summarizes the key properties of $n$-$\al\bt$  graphene-fluorographene nanoribbons.
\section{Tight-binding model}
To gain deeper insights into the DFT results and develop a flexible  modeling framework, we now construct a single-particle tight-binding model.  This approach is particularly valuable as a starting point to later incorporate the local electron-electron interactions that drives the observed magnetic ordering. We begin by identifying the orbital character of the bands near the Fermi energy ($\mt{E_{F}}$), as these states are the most relevant ones.
\begin{figure*}[t]
\includegraphics[width=.95\textwidth]{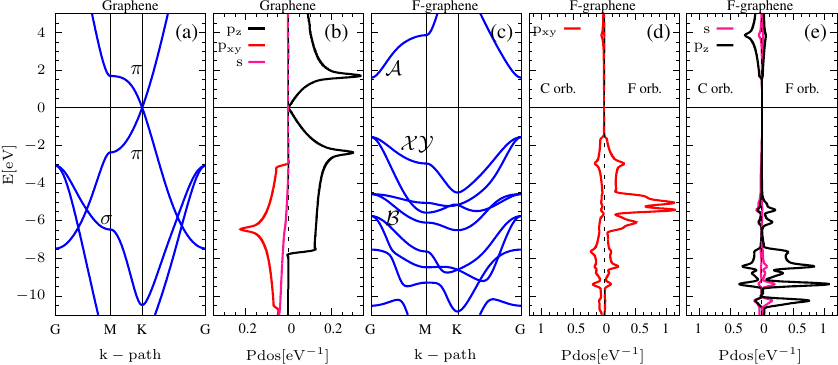}
\vspace{0cm} \caption{\label{G-FG-ES} Electronic band structure and projected density of states (PDOS) for bulk graphene and fluorographene. (a) Graphene band structure, with the $\pi$ and $\sigma$ states labeled. (b) Graphene PDOS. (c) Fluorographene band structure. (d), (e) Fluorographene PDOS projected onto F and C atoms. States labeled $\mc{A}$, $\mc{B}$, and $\mc{XY}$ in (c) will be used to describe the electronic structure of $n$-$\al\al$ and $n$-$\al\bt$ nanoribbons in the following sections.}
\end{figure*}
\subsection{Orbital projected band structure of the $\al\al$ and $\al\bt$ nanoribbons.}
To identify the orbital character of the hybridized states in the nanoribbons, we first analyze the electronic structure of bulk graphene and fluorographene, along with their projected densities of states (PDOS), as shown in Fig.~\ref{G-FG-ES}. 
Bulk graphene exhibits characteristic $\pi$ states near $\mt{E_{F}}$ with $\mt{p_{z}}$ orbital character \cite{Dresselhaus_2000}. Additionally, $\sigma$ states lie further away from $\mt{E_{F}}$ [Figs.~\ref{G-FG-ES}(a) and \ref{G-FG-ES}(b)]. In fluorographene [Figs.~\ref{G-FG-ES}(c)--\ref{G-FG-ES}(e)], we identify three key sets of states: (i)  the $\mc{A}$ states, which are above $\mt{E_{F}}$ and have a dominant $\mt{p_{z}}$ character mixed with  $\mt{s}$  states; (ii) the $\mc{XY}$ states,  which are below $\mt{E_{F}}$ and have primarily $\mt{p_{xy}}=\mt{p_{x}}+\mt{p_{y}}$ character; and (iii) the $\mc{B}$ states, which are found below $-4$~eV, and have mixed $\mt{p_{z}}$ and $\mt{s}$ character.
\begin{figure*}[t]
\includegraphics[width=.95\textwidth]{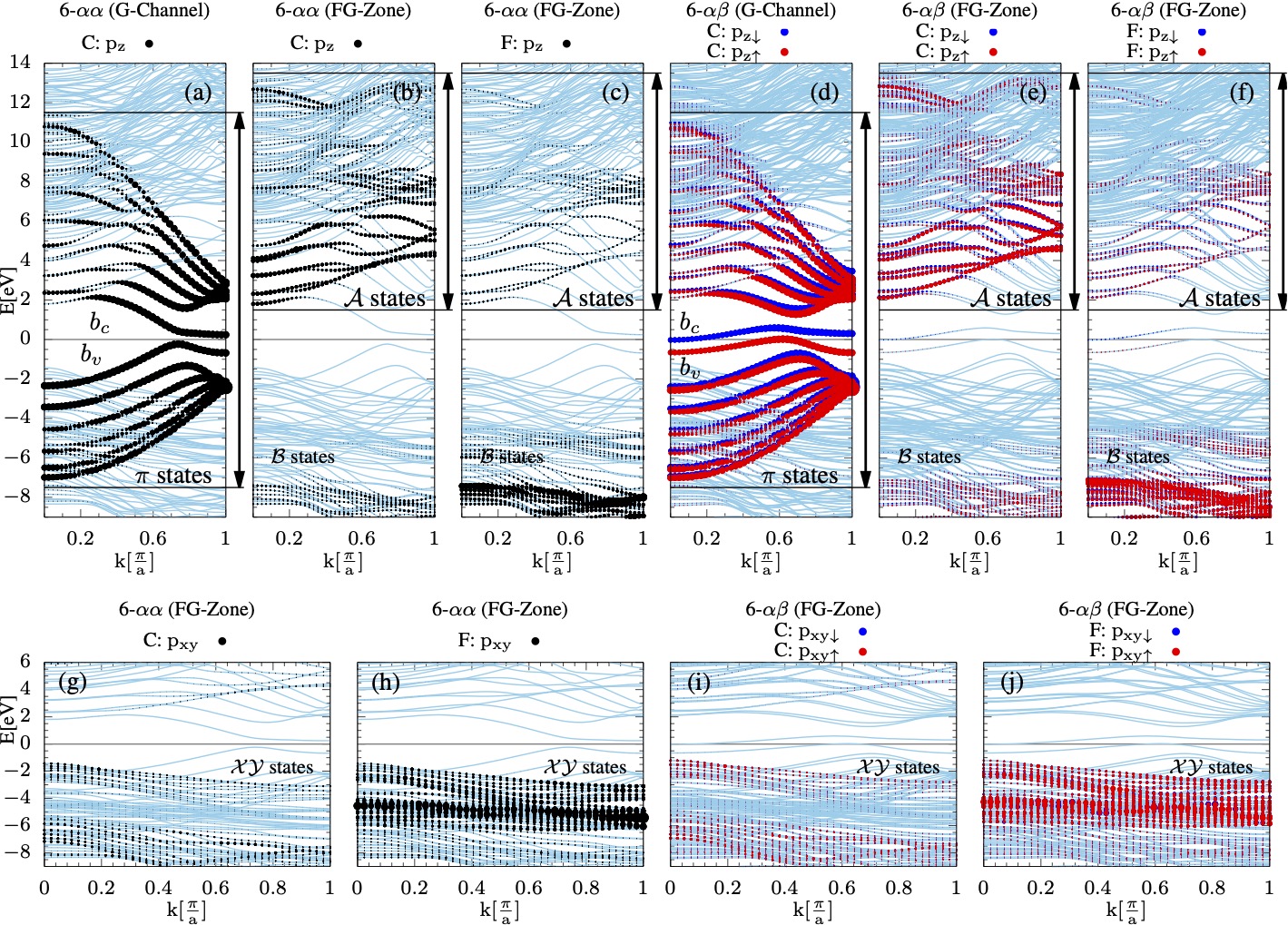}
\caption{ Band structure of the (a)-(c), (g), (h) $6$-$\al\al$  and (d)-(f), (i), (j) $6$-$\al\bt$  systems. The projection (weight) on some relevant orbitals is highlighted. The size of each ($\mt{k},\mt{E}$) point reflects the weight of the eigenstates. The bands are classified based on the dominant orbital character from the reference states in Fig.~\ref{G-FG-ES}: $\pi$ states ($\mt{p}_{z}$ character) from graphene (a), (d). 
$\mc{A}$ and $\mc{B}$ states ($\mt{p}_{z}$ character) mainly located in the fluorographene zones (b), (c), (e), (f). 
$\mc{XY}$ states ($\mt{p_{x} + p_{y}}$ character) mainly located in the fluorographene zones (g)--(j).
In the $6$-$\al\al$ system, the conduction band ($b_{c}$) near $\mt{k}\sim$0 exhibits a hybridization of $\pi$ and $\mc{A}$ states, while the valence band ($b_{v}$) is primarily composed of $\pi$ states. A similar behavior is observed for the $6$-$\al\bt$ system, where both $b_{c}$ and $b_{v}$ bands are dominated by $\pi$ states.
}
\label{QE_KPDOS}
\end{figure*}
The $n$-$\al\al$ and $n$-$\al\bt$ states result from a hybridization between some of these specific graphene and fluorographene orbitals, reflecting the superlattice structure of these systems.
Figure \ref{QE_KPDOS} presents the orbital-projected band structures of $6$-$\al\al$ and $6$-$\al\bt$ nanoribbons. The dot size reflects the weight of the eigenstates. Building upon our analysis of bulk graphene and fluorographene (Fig.~\ref{G-FG-ES}), we identify the following main features:
\begin{enumerate}[(i)]
\item Dominant $\mt{\pi}$ character: Figs. \ref{QE_KPDOS}(a) and \ref{QE_KPDOS}(d) highlight the dominance of $\mt{\pi}$ states ($\mt{p_{z}}$ orbitals from graphene carbon atoms) near $\mt{E_{F}}$, spanning the energy range $[-8$~eV, $12$~eV$]$. The valence ($b_{v}$) and conduction ($b_{c}$) bands are primarily composed of these $\mt{\pi}$ states, but exhibit crossings with bands of other orbital character.
\item Hybridization with  $\mc{A}$ states: Figs. \ref{QE_KPDOS}(a)--\ref{QE_KPDOS}(f) reveal hybridization between $\mt{\pi}$ and  $\mc{A}$ states contributing to the conduction band, with dominant C-$\mt{p_{z}}$ character.  This is evident from the states within $[1$~eV, $12$~eV$]$ exhibiting both $\mt{\pi}$ and $\mc{A}$ orbital character due to their presence in both the graphene and fluorographene regions. The spin splitting observed in the $n$-$\al\bt$ bands near $\mt{k} = 0$ (particularly in the $2$--$6$~eV window) further supports this $\mt{\pi}$-$\mc{A}$ hybridization.
\item Minimal role of $\mc{XY}$ and $\mc{B}$  states: The lower panels of Fig.~\ref{QE_KPDOS} demonstrate negligible hybridization between the $\mt{\pi}$ and $\mc{XY}$ states (dominant F-$p_{xy}$ character, located within $[-1$~eV, $-8$~eV$]$). The absence of simultaneous $\mc{XY}$ and $\mt{\pi}$ character within the $b_{v}$ band, along with the lack of spin polarization affecting $\mc{XY}$ states, establishes their minimal contribution to $b_{v}$.  Similarly, the $\mc{B}$  states (dominant F-$\mt{p_{z}}$ character, within $[-4$~eV, $-11$~eV$]$) lie well below the valence and conduction bands, making their contribution negligible.
\end{enumerate}
As a consequence, to construct a simplified electronic model reproducing the $b_{v}$ and $b_{c}$ bands, the hybridization between the $\mt{\pi}$ and $\mc{A}$ states is the crucial ingredient. 
\subsection{Spatial structure of $b_{v}$ and $b_{c}$ states}
\begin{figure}[b]
\includegraphics[width=0.9\columnwidth]{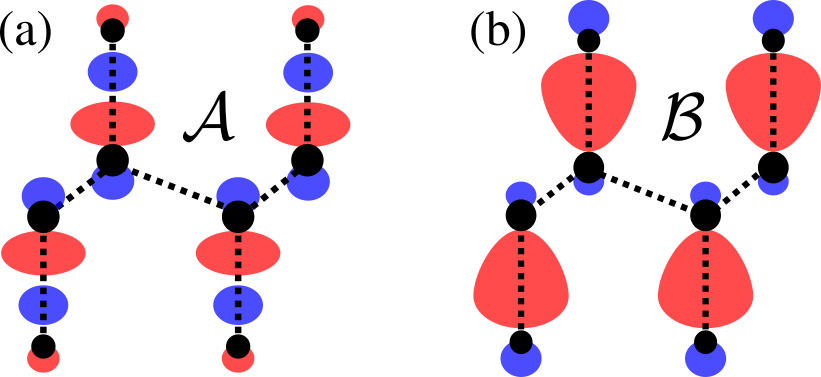}
\caption{ Schematic illustrating the formation of antibonding $\mc{A}$ and bonding $\mc{B}$ orbitals in a fluorographene channel. By combining (adding, left) and subtracting (right) atomic $\mt{sp}$ orbitals at each $\mt{C}$ and $\mt{F}$ atom, the bonding and antibonding $\mc{A}$ and $\mc{B}$ orbitals are formed. Different
colors (grayscale) represent the sign of the phase of the orbital wave function.}
\label{WAN_01}
\end{figure}

\begin{figure}[t]
\includegraphics[width=0.9\columnwidth]{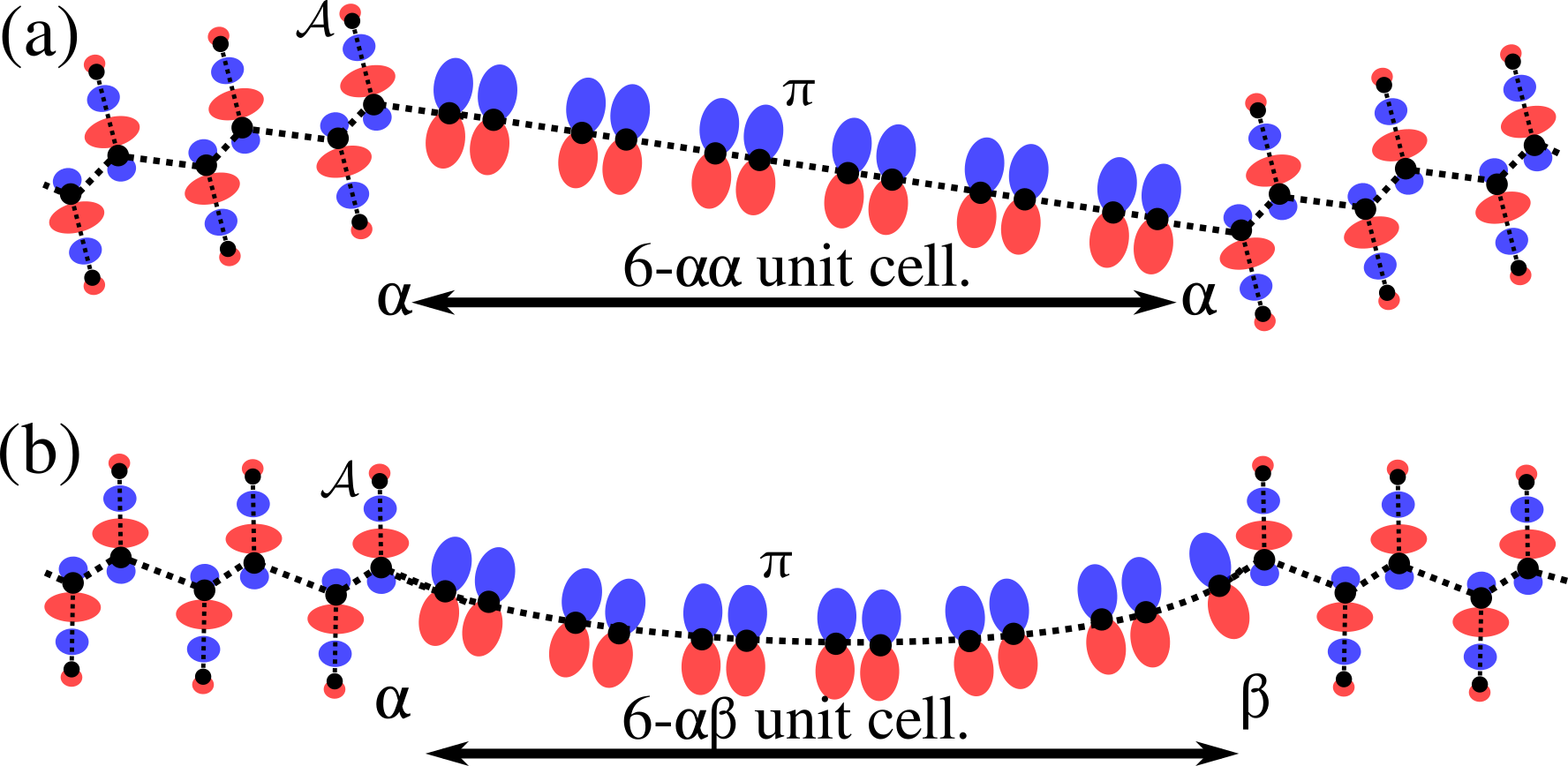}
\caption{Orbital basis of the tight-binding Hamiltonian used to simulate the electronic properties of the graphene/fluorographene nanoribbon superlattice. The basis sets for (a) 6-$\al\al$ and (b) 6-$\al\bt$ systems. Each panel depicts a unit cell containing both graphene and fluorographene regions. The basis set \{$\mt{{\pi}:{\mc{A}}}$\} comprises ${\pi}$-like orbitals ($\mt{p_{z}}$ character) from graphene channels and $\mt{{\mc{A}}}$ orbitals (primarily $\mt{p_{z}}$ character) from fluorographene.}
\label{WAN_02}
\end{figure}

We now seek a set of orbitals that will allow us to reproduce  the $b_{v}$ and $b_{c}$ bands. Our analysis showed that the $b_{v}$  states have a dominant $\mt{p_{z}}$ character and minimal hybridization with other orbitals (Fig.~\ref{QE_KPDOS}). Therefore, we can approximate them using a modulated distribution of pure $\mt{p_{z}}$ orbitals.
In contrast, the $b_{c}$ states are formed through a hybridization of $\mt{p_{z}}$ and $\mc{A}$ orbitals. As shown in Figs.~\ref{G-FG-ES} and \ref{QE_KPDOS}, the $\mc{A}$ states have significant contributions from C-$\mt{p_{z}}$, F-$\mt{p_{z}}$, C-$\mt{s}$, and F-$\mt{s}$ orbitals within the fluorographene channels. To understand the nature of these hybridized orbitals, we examine the charge distribution of $\mc{A}$ (antibonding) and $\mc{B}$ (bonding) states in fluorographene (Fig.~\ref{WAN_01}). Bonding $\mc{B}$ states have maximum charge density between atoms, while antibonding $\mc{A}$ states exhibit charge density centered on C and F atoms.  The antibonding combination, with more nodes, is higher in energy, aligning with the relative energetic positions of $\mc{A}$ and $\mc{B}$ states in Fig. \ref{G-FG-ES}(c).  The dominant C-$\mt{p_{z}}$ character in $\mc{A}$ and F-$\mt{p_{z}}$ character in $\mc{B}$ are also a consequence of the charge distribution seen in Fig. \ref{WAN_01}.
Figure \ref{WAN_02} depicts the minimal orbital basis for a tight-binding description of the $b_{v}$ and $b_{c}$ bands of $\al\al$ and $\al\bt$ systems. It comprises the $\mt{p_{z}}$ (${\pi}$) orbitals of the graphene channels and the $\mc{A}$ orbitals of fluorographene.
\subsection{Effective tight-binding Hamiltonian from WANNIER90}
Our aim is to demonstrate that the $\mt{{\pi}}$ and $\mt{{\mc{A}}}$ orbitals (defined above) are sufficient to reproduce the electronic states around $\mt{E_{F}}$ in the $\al\al$ and $\al\bt$ nanoribbons. To achieve this, we employ the Wannier90 package. This software allows us to compute maximally localized Wannier functions (MLWFs) from a set of Bloch states.  This calculation yields a transformation matrix  that  converts our basis set by combining an algebraic change of basis with an inverse Fourier transform. Furthermore, WANNIER90 generates a tight-binding Hamiltonian $\mc{H}_{_W}$ in the MLWF basis using the Kohn-Sham eigenvalues obtained from DFT. With a successful Wannierization, the eigenvalues and eigenvectors of $\mc{H}_{_W}$  exactly reproduce the DFT band structure and Kohn-Sham states within our chosen energy window. This process allows us to obtain an exact tight-binding Hamiltonian for the system.
\begin{figure}[b]
\includegraphics[width=0.9\columnwidth]{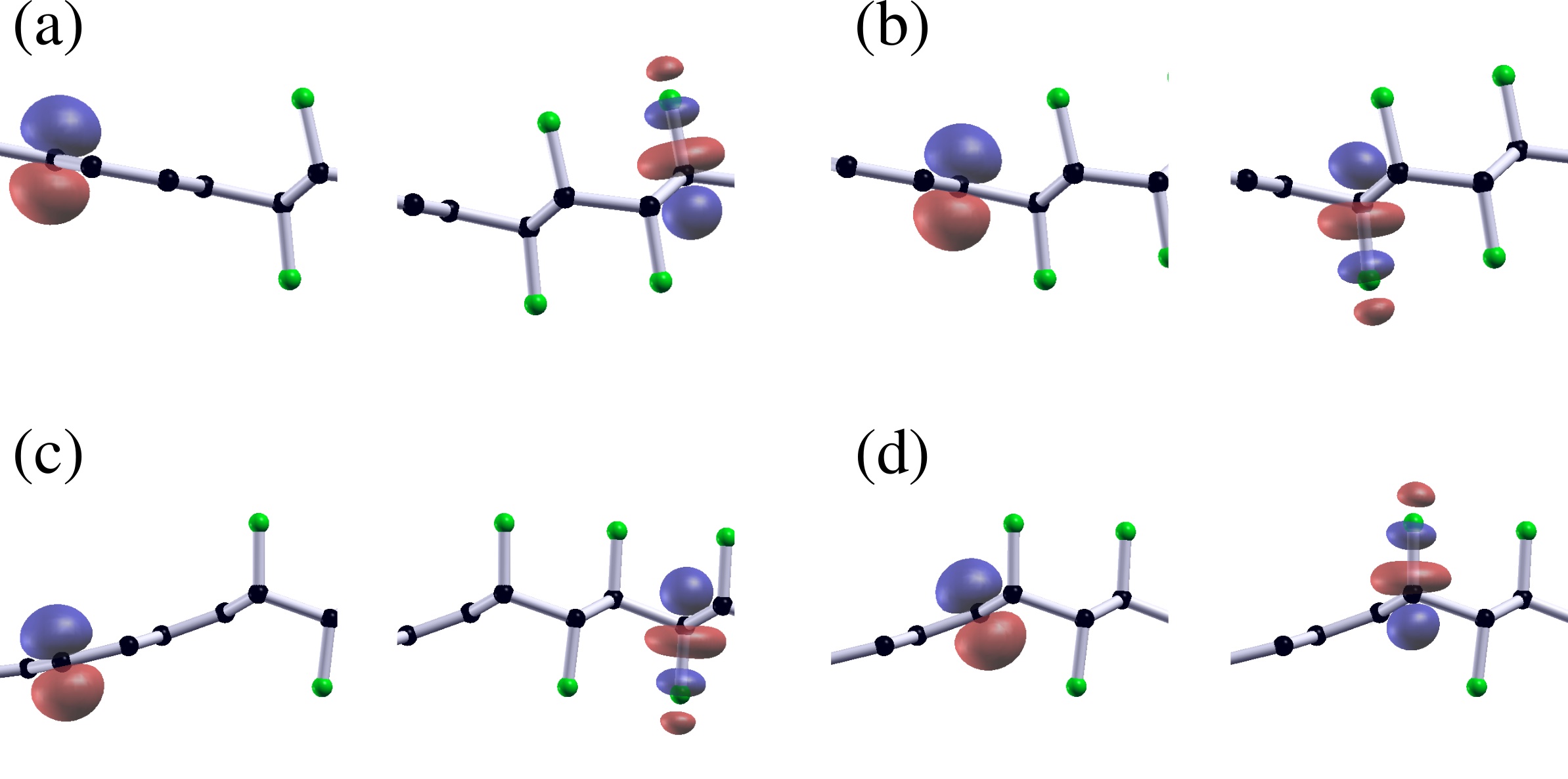}
\caption{ Maximally localized Wannier functions (MLWFs) for 6-$\al\al$ and 6-$\al\bt$ nanoribbons, calculated using Wannier90. Isosurfaces are colored (grayscale) according to the sign of the wavefunction. The obtained MLWFs closely resemble the initial guess orbitals (Fig. \ref{WAN_02}). (a), (c) MLWFs located far from the nanoribbon interfaces; (b), (d) MLWFs at the interfaces. Similar MLWF structures were obtained for nanoribbons ranging from $4$-$\al\bt$ to $8$-$\al\bt$, indicating that the MLWFs are robust against changes in system size within this range. }
\label{Wannier}
\end{figure}
In our case, a successful WANNIER90 calculation requires the following four key inputs. (i) Number of MLWFs: To reproduce the $b_{c}$ and $b_{v}$ bands, we need $12$ $\mt{{\pi}}$ and $12$ $\mt{{\mc{A}}}$ orbitals for the $6$-$\al\al$ system, and $13$ $\mt{{\pi}}$ and $11$ $\mt{{\mc{A}}}$ for the $6$-$\al\bt$ system (as seen in Fig. \ref{WAN_02}).  Thus, a total of $24$ orbitals for both cases.  
 (ii) Energy window for MLWF generation:  Based on the orbital analysis in Fig.~\ref{QE_KPDOS}, we use $[-7.5$~eV, 
 $13.5$~eV$]$ to encompass all $\mt{{\pi}}$ and $\mt{{\mc{A}}}$ states contributing to the $b_{v}$ and $b_{c}$ bands. (iii) Energy window for the tight-binding Hamiltonian: To focus on the $b_{v}$ and $b_{c}$ bands, we select the tighter window $[-1$~eV, $2.2$~eV$]$. This excludes  $\mc{XY}$ states  and higher-energy states with $\mt{C-p_{xy}}$ character (Fig.~\ref{QE_KPDOS}). (iv) Initial MLWF conditions: We approximate the desired structure using appropriate atomic  orbitals as initial seeds for each MLWF. The $\mt{{\pi}}$ orbitals are initiated as $\mt{C-p_{z}}$ orbitals centered on each carbon atom in the graphene, while $\mt{{\mc{A}}}$ as $\mt{p_{z}}$ orbitals centered between each C-F pair in fluorographene.  The Wannier calculations yielded MLWFs  that closely resemble the orbitals proposed in Fig.~\ref{WAN_01}, demonstrating the success of our approach; see Fig.~\ref{Wannier}.
\begin{figure}[t]
\includegraphics[width=0.95\columnwidth]{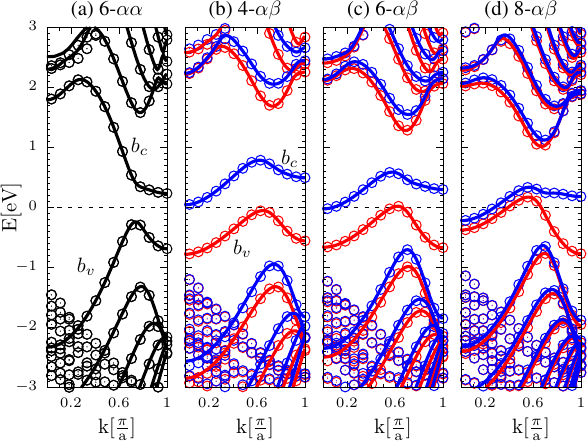}
\caption{Comparison of electronic bands obtained from DFT calculations (circles) and the spin-polarized Wannier tight-binding Hamiltonian ($\mc{H}_W^{\sg}$, lines). The $\mc{H}_W^{\sg}$ eigenvalues accurately reproduce the DFT bands around the Fermi level for all cases.  We plot only the bands derived from the ${\pi}$ and $\mt{{\mc{A}}}$ orbitals (as defined in Fig. \ref{WAN_02}).  For the $n$-$\al\bt$ systems, the red and blue lines represent the majority and minority spin bands from $\mc{H}_W^{\sg}$, respectively.  Since the $\al\al$ system is non-magnetic, the spin-up and spin-down bands coincide.}
\label{Fit_Wannier}
\end{figure}

Figures \ref{bands_AA} and \ref{bands_AB} revealed the distinct magnetic properties of the $n$-$\al\al$ and $n$-$\al\bt$  nanoribbons. The $\al\al$ systems exhibit antiferromagnetic order and are semiconducting. Conversely, $n$-$\al\bt$ systems are ferromagnetic and undergo a semiconductor-to-metal transition as their size increases. To model these magnetic properties, we extend the Wannierization process to obtain a spin-polarized tight-binding Hamiltonian, $\mc{H}_W^{\sg}$,  with $\sg=\up,\dn$.  Figure \ref{Fit_Wannier} confirms the validity of this approach by demonstrating excellent agreement between the spin-polarized $\mc{H}_W^{\sg}$ eigenvalues and the DFT bands near the Fermi level. Importantly, this comparison focuses solely on the bands derived from the previously selected  $\mt{{\pi}}$ and $\mt{{\mc{A}}}$ orbitals.

\begin{figure}[t]
\includegraphics[width=0.95\columnwidth]{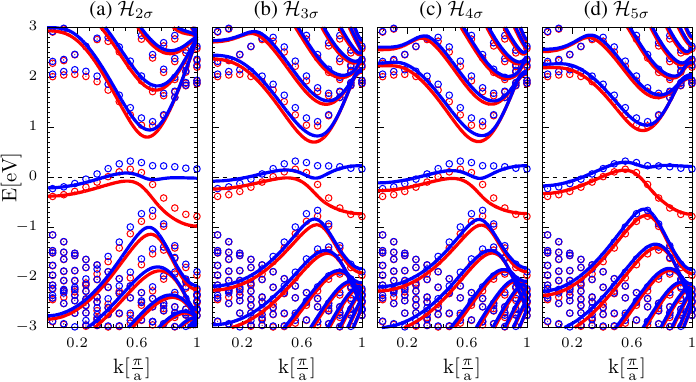}
\caption{Comparison of accuracy achieved by different simplified tight-binding Hamiltonians ($\mc{H}_{j\sg}$) in describing the DFT results for the $8$-$\al\bt$ nanoribbon. $\mc{H}_{j\sg}$ refers to a Hamiltonian where matrix elements beyond the $j$th nearest neighbor are excluded. Circles represent the DFT data and lines represent the $\mc{H}_{j\sg}$ results. The figure demonstrates that using $\mc{H}_{5\sg}$ is necessary to achieve excellent agreement with the exact DFT results in the $\al\bt$ case. Red and blue represent
the majority and minority spin bands, respectively.}
\label{H_Wannier}
\end{figure}

The Wannier-based tight-binding Hamiltonian, while exact, contains many parameters. A crucial question for constructing a simplified Anderson-Hubbard model (see next section) is to analyze how to reduce the parameter set while retaining accuracy. Figure \ref{H_Wannier} addresses this question for the $8$-$\al\bt$ nanoribbon. We construct various $\mc{H}_{j\sg}$ Hamiltonians, progressively including interactions up to the fifth-nearest neighbor, and compare them to the DFT results. We label these Hamiltonians $\mc{H}_{j\sg}$, where $j$  indicates the farthest-neighbor interactions included. It is apparent from the figure that $\mc{H}_{5\sg}$ provides an excellent overall fit to the DFT bands.  Importantly, $\mc{H}_{j\sg}$ Hamiltonians with $j<5$ fail to reproduce the semiconductor-to-metal transition observed in  $\al\bt$ systems (Fig.~\ref{bands_AB}), highlighting the need for up to the fifth-nearest-neighbor interactions to accurately capture this behavior.
While $\al\al$ systems can be reasonably modeled with fewer parameters (second-nearest neighbors), these results underscore the importance of analyzing a larger number of nearest-neighbors to ensure our simplified models capture the key physics of the $\al\bt$ systems.
\section{Mean field Anderson-Hubbard Model from Wannier calculations}
Our Anderson-Hubbard model ($H_{AH}$) is constructed using a basis of Wannier orbitals $\{ \mt{{\pi}},\mt{{\mc{A}}}\}$,
\begin{equation}
H_{AH} \,=\, \sum_{j,\sg} \ve_{j}\,c_{j\sg}^{\dag }\,c_{j\sg}^{{}}-\sum_{j,l,\sg} t_{jl}\,c_{j\sg}^{\dag }c_{l\sg} \,+\,\sum_{j}U_{j}\, \hat{n}_{j\up} \, \hat{n}_{j\dn} \,,
\label{AH_01}
\end{equation}
where the operator $c^{\dag}_{j\sg}$ ($c_{j\sg}$) creates (annihilates) an electron with spin $\sg$ at site $j$ (in the corresponding orbital) and $\hat{n}_{j\sg}=c^{\dag}_{j\sg} c_{j\sg}$ is the number operator. Here, $\ve_{j}$ is the local energy of site $j$,  $t_{jl}$ is the matrix element between $j$ and $l$ sites, and  $U_{j}$ is the Hubbard constant at site  $j$. 

We solve $H_{AH}$ in a mean-field description by using the Hartree-Fock method, so the interacting term $\hat{n}_{j\up} \, \hat{n}_{j\dn}$  is substituted by $\la \hat{n}_{j\up} \ra \, \hat{n}_{j\dn} \,+\, \hat{n}_{j\up} \, \la  \hat{n}_{j\dn} \ra $. Thus,
\begin{equation}
H_{AH}\xrightarrow{} \bar{H}_{AH}\, = \sum_{j,\sigma} {\ve_{j\sg}  \,c_{j\sg}^{\dag} c_{j\sg} -\sum_{j,l,\sg} t_{jl}\,c_{j\sg}^{\dag}c_{l\sg} }\,,
\label{AH_02}
\end{equation}
where $\ve_{j\sg}$ is the local energy of spin $\sg$ that depends of the occupation $\la \hat{n}_{j\bar{\sg}} \ra$ of the opposite spin $\bar{\sg}$ at the $j$ site, namely,
\begin{eqnarray}
\nonumber
\ve_{j\up} &=& \ve_{j} + U_{j} \la \hat{n}_{j\dn} \ra \\
\ve_{j\dn} &=& \ve_{j} + U_{j} \la \hat{n}_{j\up} \ra\,.
\label{MF_HB}
\end{eqnarray}
Equation \eqref{MF_HB} can be used to obtain $U_{j}$, that is,
\begin{equation}
U_{j} = - \frac{ \ve_{j\up} - \ve_{j\dn} }{ \la \hat{n}_{j\up} \ra - \la \hat{n}_{j\dn} \ra } = -\frac{\Delta \ve_{j\sg}}{ \la \hat{m}_{j} \ra}\,,
\label{AH_03}
\end{equation}
with $ \la \hat{n}_{j\sg} \ra = \sum_{\ve_{(\mt{k},\sg)} \le \mt{E_{F}}} |\psi_{(\mt{k,\sg})\,l}|^{2}$.
Thus, from $\ve_{j\sg}$, $\ve_{(\mt{k},\sg)}$, and $\psi_{(\mt{k,\sg})}$, which are the site energy, eigenvalues, and eigenvectors of $\mc{H}_W^{\sg}$, respectively, one can get an estimate of $U_{j}$ for those sites with nonzero spin polarization. Similarly, the local energy $\ve_{j}$ can also be obtained through Eq. \eqref{MF_HB}, after calculating $U_j$,
\begin{equation}
\ve_{j} = \frac{1}{2} \left( \ve_{j\up} + \ve_{j\dn} - \la \hat{n}_{j} \ra \, U_{j} \right)\,,
\label{AH_05}
\end{equation}
where $\la \hat{n}_{j}\ra =\la \hat{n}_{j\up} \ra + \la \hat{n}_{j\dn} \ra  $ is the total electronic occupation at site $j$. A detailed procedure to obtain all the necessary Anderson-Hubbard model parameters is described in the Appendix.

After obtaining the necessary parameters from the analysis of the  $6$-$\al\al$ and $8$-$\al\bt$ systems, we now investigate the band structure and magnetic properties of other $n$-$\al\al$ and $n$-$\al\bt$ nanoribbons using the AH model. We first validate our model by contrasting its predictions against DFT calculations for systems up to $10$-$\al\al$ and $10$-$\al\bt$.  After this crucial step, we extrapolate the AH model and explore larger systems inaccessible to DFT calculations. 
\begin{figure}[t]
\includegraphics[width=0.95\columnwidth]{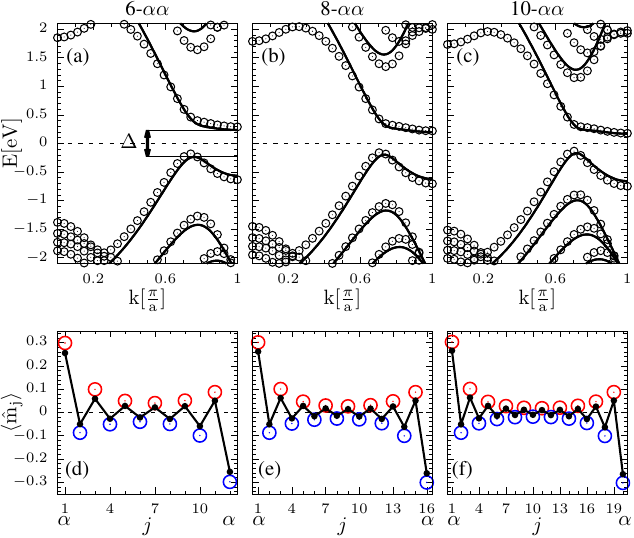}
\caption{ Validation of the mean-field AH model for $n$-$\al\al$ systems.  Upper panels: Comparison of electronic band structures obtained from DFT (open circles) and the AH model (lines) using $\bar{H}_{AH}$  with up to second-nearest-neighbor interactions. Results are shown for (a) $6$-$\al\al$, (b) $8$-$\al\al$, and (c) $10$-$\al\al$ systems.  Lower panels: Corresponding spin polarization along the graphene channel, calculated using DFT projected density of states (PDOS, red and blue open circles) and $\bar{H}_{AH}$ eigenvectors (black filled circles). Excellent agreement across all systems validates the AH model.}
\label{AH_aa_1}
\end{figure}

Figure \ref{AH_aa_1} demonstrates the AH model's accuracy for $6$-$\al\al$, $8$-$\al\al$, and $10$-$\al\al$ nanoribbons.  
The band structures calculated with the AH model (lines), including up to second-nearest-neighbor hopping terms, fairly reproduce the main features of those obtained from DFT (open circles). The AH model reproduces the semiconducting behavior and the band gaps, thus confirming its validity for these systems.
While agreement near the Fermi level is excellent, discrepancies arise closer to the Brillouin zone center ($\mt{k}=0$). This is to be expected, even for $6$-$\al\al$, due to our simplified model including only up to second-nearest-neighbor interactions. The inclusion of hopping terms up to sixth-nearest neighbors significantly improves the agreement across the entire Brillouin zone and beyond the Fermi level (not shown) \cite{Guzman_thesis}.

\begin{figure}[t]
\includegraphics[width=0.95\columnwidth]{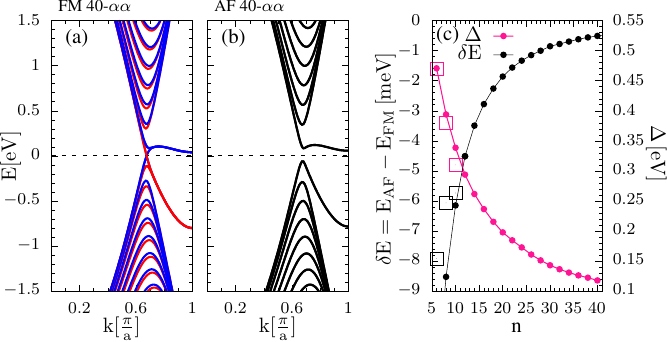}
\caption{ Band structure and magnetic properties of the  $n$-$\al\al$ systems, calculated within the mean-field AH model. (a) Band structure for a $40$-$\al\al$ nanoribbon exhibiting a metastable ferromagnetic (FM) state.  (b) Band structure for the antiferromagnetic (AF) ground state. (c) Energy difference between AF and FM states, showing the stability of the AF phase and the evolution of its energy gap as a function of $n$. DFT results are shown for comparison (squares). }
\label{AH_aa_2}
\end{figure}

We now explore system sizes that are inaccessible to DFT. Figure \ref{AH_aa_2} analyzes a $40$-$\al\al$ nanoribbon---recall from Fig.~\ref{SP_AA} that the $n$-$\al\al$ systems exhibit an antiferromagnetic (AF) ground state, with opposite spin alignment across the nanoribbon.
%
Figure \ref{AH_aa_2}(a) depicts the band structure in the metastable ferromagnetic (FM) state, a revealing metallic behavior. In contrast, the AF ground state [Fig.~\ref{AH_aa_2}(b)] is semiconducting. Crucially, Fig.~\ref{AH_aa_2}(c) demonstrates that the AF state is energetically favorable across the studied system sizes (up to $40$-$\al\al$), although the energy difference ($|\delta E|=|E_{AF}-E_{FM}|$) decreases with increasing size.  Figure \ref{AH_aa_2}(c) further illustrates how the AF bandgap ($\Delta$) also decreases with increasing $n$.
\begin{figure}[b]
\includegraphics[width=0.95\columnwidth]{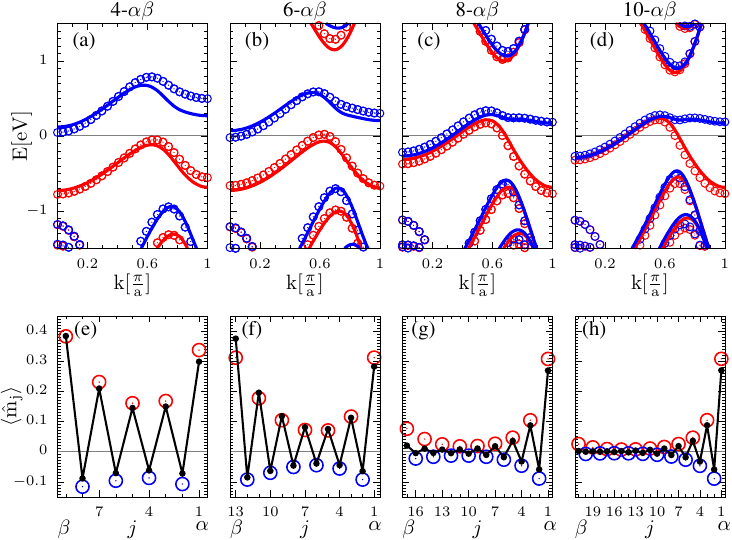}
\caption{ Comparison of DFT spin-polarized band structures (red and blue circles) with the mean-field AH model predictions (red and blue lines) for $n$-$\al\bt$ systems ($n=4,6,8,10$).  The AH model accurately captures the key electronic features across all system sizes, validating our approach. Lower panels: corresponding spin polarization along the graphene channel, similar to Fig. \ref{AH_aa_1}.}
\label{AH_ab_1}
\end{figure}

We now consider the $n$-$\al\bt$ nanoribbons. As before, we first validate the AH model  by comparing it to  DFT calculations for  $4$-$\al\bt$ through $10$-$\al\bt$ systems (Fig.~\ref{AH_ab_1}).  We include up to the fifth-nearest-neighbor hopping terms based on the $8$-$\al\bt$ case (see the Appendix).
Figure \ref{AH_ab_1} demonstrates an excellent agreement between AH and DFT band structures, specially for large system sizes. Notably, the AH model accurately captures the semiconductor-to-metal transition with increasing system size.
The AH model also reproduces the global ferromagnetic state and local spin behavior (lower panels of Fig.~\ref{AH_ab_1})---spin polarization oscillates along the channel, as found in DFT calculations. Furthermore, both methods show decreasing spin polarization near the $\bt$ interface with increasing size, converging towards zero magnetic moment at the $\bt$ edge in $10$-$\al\bt$.  Conversely, the spin polarization at the $\al$ edge remains constant, a trend that persists in larger systems (we verified it up to $n=40$).

\begin{figure}[t]
\includegraphics[width=0.95\columnwidth]{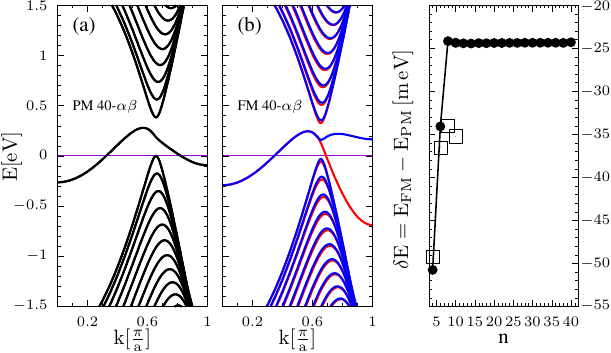}
\caption{Analysis of magnetic states in larger $\al\bt$ systems. (a) Band structure for a $40$-$\al\bt$ nanoribbon in the paramagnetic (PM) state. (b) Band structure in the ferromagnetic (FM) state. (c) Energy difference per unit cell ($\delta E=E_{FM}-E_{PM}$) as a function of the number of zigzag chains ($n$), up to $n = 40$. DFT results are shown for comparison (squares). As with smaller systems, spin polarization in (b) is localized around the $\al$ interface.}
\label{AH_ab_2}
\end{figure}
Figure \ref{AH_ab_2} compares the paramagnetic (PM) and ferromagnetic states of the $40$-$\al\bt$ system.  In the PM state, [Fig. \ref{AH_ab_2}(a)], electronic states near  $\mt{k}=\frac{\pi}{a}$ are fully occupied. In contrast, the FM state, [Fig. \ref{AH_ab_2} (b)], exhibits spin splitting, with one spin band filled and the other empty near $\mt{k}=\frac{\pi}{a}$.  This difference reflects the presence of spin polarization along the $\al$ edge in the latter. Finally, Fig.~\ref{AH_ab_2}(c) reveals a sharp increase in $\delta E$ for small system sizes, attributed to the semiconductor-metal transition depicted in Fig. ~\ref{bands_AB}. For larger systems, the energy difference ($\delta E$) between the FM and PM states remains nearly constant. This is due to the localized nature of the edge-state interactions responsible for the magnetic ordering.
\section{Summary}
In this work, we studied the electronic and magnetic properties of graphene channels embedded within fluorographene, focusing on two distinct interfaces: the fully fluorinated zigzag chain ($\al$) and the half-fluorinated zigzag chain ($\bt$).  

In the case of the $n$-$\al\al$ channels, we found that they exhibit similarities to pristine zigzag graphene nanoribbons: they display antiferromagnetic ordering, semiconducting behavior, and a decreasing energy gap with increasing channel width.  The $n$-$\al\bt$ systems, on the other hand, are ferromagnetic and undergo a semiconductor-to-metal transition with increasing channel width.  In the semiconducting phase,  spin polarization is present at both edges.  For wider channels,  dominant spin polarization persists only at the $\al$ edge, with vanishing polarization at the $\bt$ edge. 

Our results demonstrate that selective fluorination of graphene edges offers an opportunity for tunable electronic and magnetic properties of significant interest for spintronic applications. This strategy provides a pathway for engineering semiconductor properties without altering the graphene channel width itself. For example, inducing metallicity in graphene nanoribbons has been a subject of interest recently \cite{rizzo:2020}; our results show that by tuning the interface and by fabricating $\al\bt$ nanoribbons, one can also obtain metallic nanoribbons.
Graphene nanoribbons are promising devices with potential uses in microelectronics, spintronics, quantum computing,  and topological electronics. Our findings highlight the potential of selective fluorination for tailoring graphene's properties, paving the way for such developments. 
\section{Acknowledgments}
We acknowledge ﬁnancial support from the ANPCyT-FONCyT (Argentina) under Grants No. PICT 2018-1509 and No. PICT 2019-0371, from CONICET Grant No. PIP 2022-2024 
(11220210100041CO), and SeCyT-UNCuyo grant 06/C053-T1. G.U. thanks Nathan Goldman for his hospitality at the Universit\'e Libre de Bruxelles (ULB).
\appendix*
\section{Parameters used in the Anderson-Hubbard model}
\begin{figure}[t]
\includegraphics[width=0.95\columnwidth]{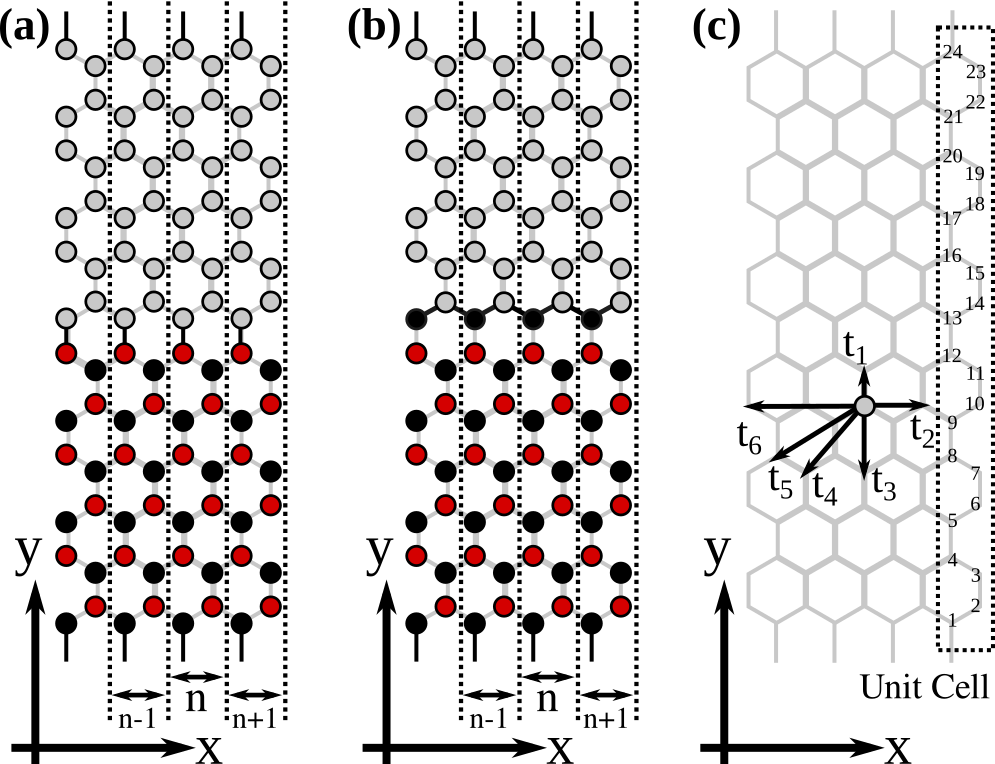}
\caption{ Crystal structure and Anderson-Hubbard model for the $6$-$\al\al$ and the $6$-$\al\bt$ nanoribbons. (a), (b) The key structural difference: The number of graphene sites (black and red circles) within the unit cell. The $6$-$\al\al$ system has $12$ sites, whereas the $6$-$\al\bt$  system has $13$. This variation significantly alters the band structures, as discussed previously. Gray circles correspond with the fluorographene sites. (c) A  simplified lattice representation of the Hamiltonian [Eq. \ref{AH_02}]. The hexagonal lattice shows the unit cell (enclosed by dashed lines) and numbered atomic sites. The model considers one orbital per lattice vertex and periodic boundary conditions in both directions.  Additionally, it includes hopping terms (notated as  $t_{n}$ with $n=1,2,3,4,5,6$) up to the sixth-nearest neighbors. }
\label{AH_MODEL}
\end{figure}

We describe here the procedure we use to obtain the different parameters of the mean field Anderson-Hubbard model described in the main text and schematically depicted in Fig.~\ref{AH_MODEL}.  Figures.~\ref{AH_MODEL}(a) and \ref{AH_MODEL}(b) highlight the locations of the ${\pi}$ (black and red dots) and the ${\mc{A}}$ (gray dots) orbitals, used for describing the  graphene and fluorographene regions, respectively. Figure \ref{AH_MODEL}(c) shows the underlying hexagonal lattice containing  one Wannier orbital per vertex. The dashed lines defines the unit cell. Periodic boundary conditions and translation invariance are assumed along the $y$ and $x$ axes, respectively.
Notice that although both $\al\al$ and $\al\bt$ systems have $24$ sites per unit cell, the distribution of ${\pi}$ and ${\mc{A}}$ orbitals at the graphene-fluorographene interface differs. The graphene region contains $12$ sites in the  $\al\al$ case and $13$ sites in the $\al\bt$ case.

\begin{figure}[b]
\includegraphics[width=0.95\columnwidth]{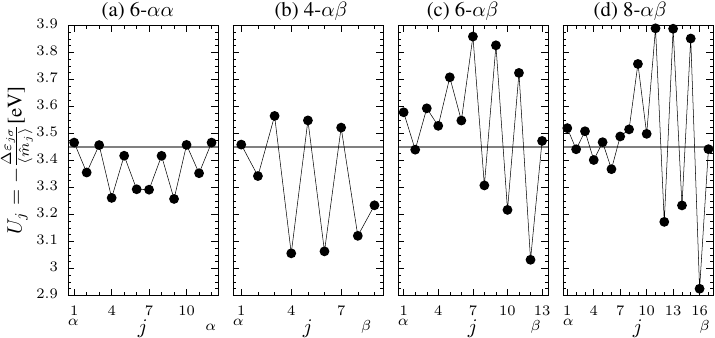}
\caption{Hubbard constant ($\mt{U_{j}}$) calculated along the graphene channels for (a) $6$-$\al\al$ , (b) $4$-$\al\bt$, (c) $6$-$\al\bt$, and (d) $8$-$\al\bt$ nanoribbons. $\mt{U_{j}}$ is obtained using Eq. (\ref{AH_03}) only at sites with spin polarization (i.e., where $\la \hat{n}_{j\up} \ra - \la \hat{n}_{j\dn} \ra \ne 0 $). The location of the $\al$  and $\bt$ edges is indicated.}
\label{AH_Hub_00}
\end{figure}

The Hubbard interaction strength $U_j$ is self-consistently calculated from  the local energies, eigenvalues, and eigenvectors [Eq.~\eqref{AH_03}] of the spin-resolved Wannier Hamiltonians ($\mc{H}_W^{\sg}$). Figure \ref{AH_Hub_00} shows its  spatial variation along the graphene channels. To simplify our model, we adopt a representative value of $U_{g}=3.45$~eV for all graphene sites as this is roughly the value obtained near the highly spin-polarized edges---except a slight deviation at the $\bt$ edge in the smallest ($4$-$\al\bt$) system.
%
%
In addition, we calculate $\mt{U}_{fg}$ at spin-polarized fluorographene edge sites and their first-nearest neighbors, differentiating between $\al$ and $\bt$ interfaces: $\mt{U}^{\al}_{fg}=5.2$~eV and  $\mt{U}^{\bt}_{fg}=2.3$~eV. The remaining fluorographene sites can be assigned $U_{g}$, as this value is less critical for our purposes since, in the energy window we want to model, there is a  negligible electronic occupation in the middle of the fluorographene channels.

\begin{figure}[t]
\includegraphics[width=0.95\columnwidth]{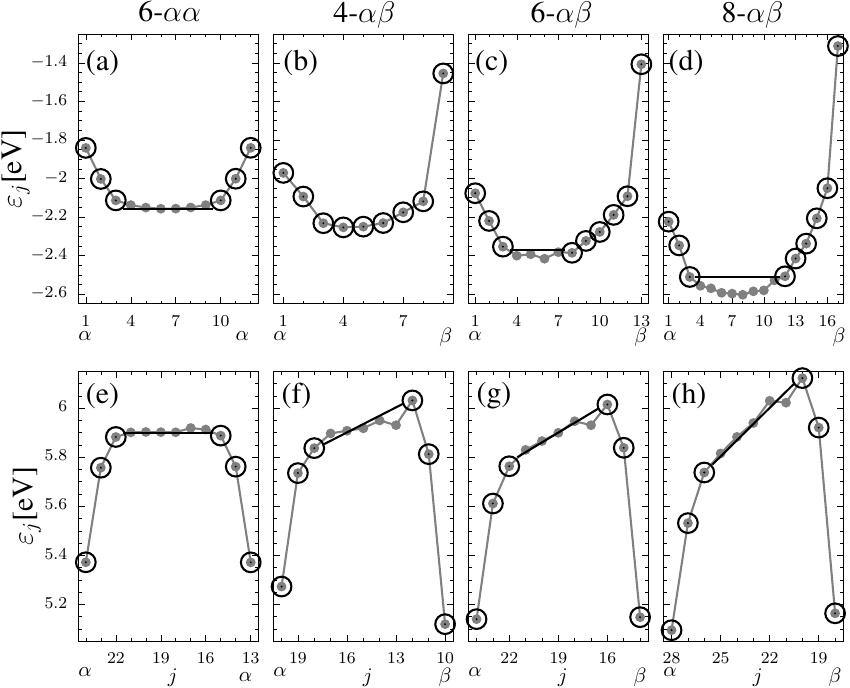}
\caption{Local energy ($\ve_{j}$) of the system calculated using Eq. (\ref{AH_03}). (a)-(d) The graphene channels and (e)-(h) the behavior along the fluorographene region. Gray circles represent $\ve_{j}$ values, which exhibit  similar behavior near $\al$ and $\bt$ edges across different nanoribbon sizes. This allows us to reliably use these values to model wider systems that are inaccessible to DFT calculations. For larger systems, $\ve_{j}$ within the central region can be described by interpolating the straight lines and employ the values at the edges (black open circles).}
\label{AH_Hub_01}
\end{figure}

The site energies $\ve_{j}$, obtained by fitting Eq.~\eqref{AH_05}, are shown in Fig.~\ref{AH_Hub_01}. In the graphene channels (upper panels), the interface effect extends approximately three sites away from $\al$ edges and up to six sites from $\bt$ edges.  The potential decreases towards the channel center.  Conversely, within fluorographene regions (lower panels of Fig.~\ref{AH_Hub_01}), the potential increases towards the center, with the interface effect extending roughly three sites from both $\al$ and $\bt$ edges. Importantly, Fig.~\ref{AH_Hub_01} reveals that the  behavior of $\ve_{j}$ near edges is consistent across different nanoribbon sizes. This allows us to model wider systems (intractable for DFT) by employing these values. For the central regions of larger systems, we interpolate $\ve_{j}$ based on the linear trends observed in our smaller systems.

Our next step is to calculate the hopping matrix elements.  In Fig.~\ref{H_Wannier}, we show the importance of including hopping terms up to second-nearest neighbors for $\al\al$ interfaces and fifth-nearest neighbors for $\al\bt$ interfaces to achieve DFT-level accuracy near the Fermi level. This means that in $\al\al$ systems, hopping matrix elements will link orbitals in adjacent unit cells, while in $\al\bt$ systems they might connect orbitals across second-nearest neighbor unit cells.
To calculate these hopping matrix elements we used the Wannier Hamiltonians for the $6$-$\al\al$  the $8$-$\al\bt$ nanoribbons. 

The hopping matrix elements are determined by 
\begin{equation}
t_{ij} = \frac{1}{2} \la {\zeta_{i},i}|\mc{H}_W^{\up} +  \mc{H}_W^{\dn}| {\zeta_{j},j} \ra\,,
\label{AH_t1}
\end{equation}
where $\zeta_{l}$ represents the $\pi$ or $\mc{A}$ orbital  depending on the site. The spin averaging eliminates a weak spin dependence of matrix elements. 
Let us first consider the  matrix elements for sites within the graphene ($t$) and fluorographene ($\tau$) channels, away from interfaces. 
Since the system is slightly anisotropic, with the goal of reducing the number of parameters, we define the average $m$th-order matrix element at each site, say  $l$, as
\begin{eqnarray}
\nonumber
\la ^{m}t_{l} \ra &=& \frac{1}{Z_{ml}}\sum_{\la i,l \ra_{m}} \, t_{il} \,,\\
\la ^{m}\tau_{l} \ra &=& \frac{1}{Z_{ml}} \sum_{\la i,l \ra_{m}} \, \tau_{il} \,,
\label{AH_t0}
\end{eqnarray}
where $Z_{ml}$ is the number of $m$th-nearest neighbors of  site $l$ and the sum on $i$ runs on those neighbors.

Table \ref{mean_hop_tab} presents those average values. Variations across different systems are small (up to $10$\%), highlighting that these values are primarily determined by the local structure. However, sites with neighbors at the interface exhibit significant changes (up to $50$\%), demonstrating the presence  of interface effects.
%
\begin{table}[t]
\begin{center}
\begin{tabular}{|c|c|c|c|c|c|c|}
\hline \hline
GC &$\la^{1}\mt{t}_{l}\ra$& $\la^{2}\mt{t}_{l}\ra$&$\la^{3}\mt{t}_{l}\ra$&$\la^{4}\mt{t}_{l}\ra$&$\la^{5}\mt{t}_{l}\ra$&$\la^{6}\mt{t}_{l}\ra$ \\ \hline
6-$\al\al$ &-2.789 & 0.237 & -0.241 & 0.022 & 0.046 &-0.020 \\ \hline 
4-$\al\bt$ &-2.767& 0.231&-0.235& 0.023& 0.046&-0.019 \\ \hline 
6-$\al\bt$ &-2.777 & 0.237 &-0.245 & 0.024 & 0.048 & -0.020 \\ \hline 
8-$\al\bt$ & -2.778 & 0.238 &-0.244 & 0.025 & 0.048 &-0.020 \\ \hline \hline
FG &$\la^{1}\mt{\tau}_{l}\ra$& $\la^{2}\mt{\tau}_{l}\ra$&$\la^{3}\mt{\tau}_{l}\ra$&$\la^{4}\mt{\tau}_{l}\ra$&$\la^{5}\mt{\tau}_{l}\ra$&$\la^{6}\mt{\tau}_{l}\ra$ \\ \hline
6-$\al\al$ &-1.425 & 0.015 &-0.172 &-0.075 &-0.002 &-0.012\\ \hline 
4-$\al\bt$ &-1.430& 0.013& -0.172& -0.073& -0.002& -0.012\\ \hline 
6-$\al\bt$ &-1.440 & 0.009 &-0.168 &-0.075& 0.001&-0.011\\ \hline 
8-$\al\bt$ &-1.431 & 0.012 &-0.176 &-0.072 &-0.003 &-0.013\\ \hline \hline
\end{tabular}
\end{center}
\caption{Average $n$th-order matrix elements at the center of graphene (GC, $\la ^{n}t_{l} \ra$) and fluorographene (FG, $\la ^{n}\tau_{l} \ra$ ) channels, calculated using Eq. \eqref{AH_t0}.}
\label{mean_hop_tab}
\end{table}
%
As expected, $\la ^{1}t_{l} \ra$ and $\la ^{1}\tau_{l} \ra$  have the largest magnitudes, as they encode the essential information about the crystal structure. In particular,  their values are significantly altered by the presence of interfaces, as discussed below. 

\begin{figure*}[t]
\includegraphics[width=0.95\textwidth]{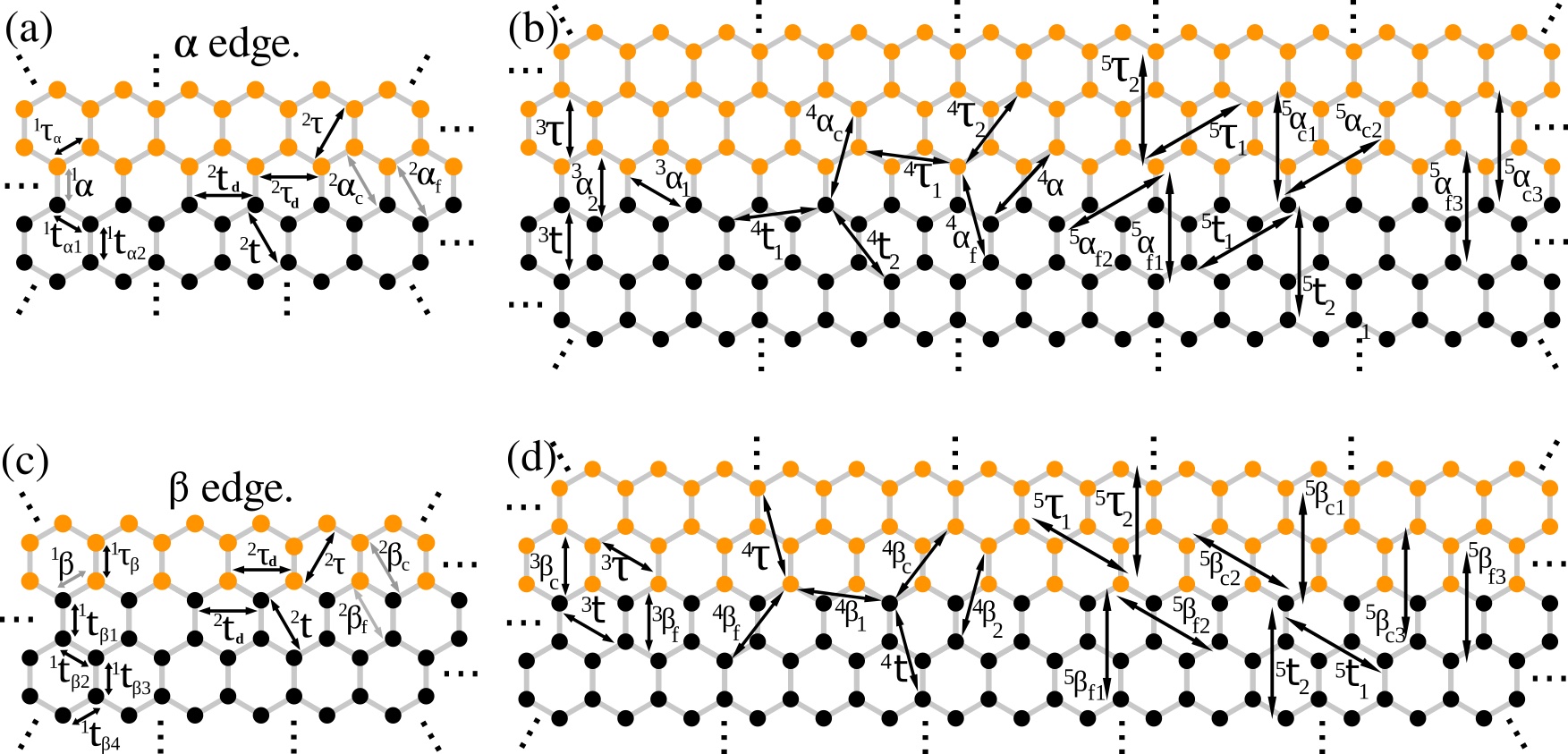}
\caption{Matrix elements between interface-adjacent orbitals in graphene (black dots) and fluorographene (orange dots).  (a), (b) The $\al$ interface;  (c), (d) The $\bt$ interface. (a), (c) First- and second-nearest-neighbor interactions. (b), (d): Third- through fifth-nearest-neighbor interactions.}
\label{Int_hop}
\end{figure*}

On the other hand, as mentioned in the main text, a successful model requires inclusion of the matrix elements beyond first-nearest neighbors to accurately capture interface effects.
Figure \ref{Int_hop} depicts the relevant matrix elements near the $\al$ and $\bt$ interfaces. Black dots represent graphene $\pi$ orbitals, while gray dots represent fluorographene $\mc{A}$ orbitals. Figures \ref{Int_hop}(a) and \ref{Int_hop}(c) show first-  and second-nearest-neighbor interactions for $\al$ and $\bt$ interfaces, respectively, Panels \ref{Int_hop}(b) and \ref{Int_hop}(d) display third- through fifth-nearest neighbor interactions, while  
\begin{table}[b]
\begin{center}
\begin{tabular}{|c|c|c|c|c|c|}
\hline \hline
$\mt{^{1}\tau_{\bt}}$&$\mt{^{1}\bt}$&$\mt{^{1}t_{\bt1}}$ &$\mt{^{1}t_{\bt2}}$&$\mt{^{1}t_{\bt3}}$&$\mt{^{1}t_{\bt4}}$\\ \hline
-1.322 & 2.174 &-2.853&-2.691& -2.894&-2.744 \\ \hline \hline \hline
$\mt{^{2}\tau}$&$\mt{^{2}\tau_{d}}$&$\mt{^{2}\bt_c}$ &$\mt{^{2}{\bt_f}}$&$\mt{^{2}t_{d}}$&$\mt{^{2}t}$\\ \hline
 0.004& 0.020& -0.043&0.033& 0.439& 0.234\\ \hline \hline
$\mt{^{3}\tau}$ &$\mt{^{3}\bt_{f}}$ &$\mt{^{3}\bt_{c}}$ &$\mt{^{3}t}$ & - & - \\ \hline
-0.142&0.322&0.209&-0.157& - & -\\ \hline \hline
$\mt{^{4}\tau}$ &$\mt{^{4}\bt_{f}}$ &$\mt{^{4}\bt_{1}}$&$\mt{^{4}\bt_{2}}$&$\mt{^{4}\bt_{c}}$ & $\mt{^{4}t}$  \\ \hline
-0.091&-0.020&0.004&-0.109&0.136&0.026\\ \hline \hline
$\mt{^{5}\tau_{2}}$ &$\mt{^{5}\tau_{1}}$&$\mt{^{5}\bt_{f3}}$&$\mt{^{5}\bt_{f2}}$&$\mt{^{5}\bt_{f1}}$ &- \\ \hline
0.012&-0.005&-0.067&-0.025&-0.057& - \\ \hline \hline
$\mt{^{5}t_{2}}$ &$\mt{^{5}t_{1}}$&$\mt{^{5}\bt_{c3}}$&$\mt{^{5}\bt_{c2}}$&$\mt{^{5}\bt_{c1}}$ &- \\ \hline
0.056&0.046&-0.017&-0.025&0.010& - \\ \hline \hline
\end{tabular}
\end{center}
\caption{Matrix elements of Figs. \ref{Int_hop}(c) and \ref{Int_hop}(d) near the $\bt$ interface, as extracted from Wannier analysis of the 8-$\al\bt$ system.}
\label{beta_int}
\end{table}
Tables \ref{alpha_int} and \ref{beta_int} contain their values. 
\begin{table}[b]
\begin{center}
\begin{tabular}{|c|c|c|c|c|c|c|}
\hline \hline
$\mt{^{1}\tau_{\al}}$&$^{1}{\al}$&$\mt{^{1}t_{\al1}}$ &$\mt{^{1}t_{\al2}}$&-&-&-\\ \hline
-1.428 & 2.348 &-2.701& -2.741 & - & - & - \\ \hline
$\mt{^{2}\tau}$&$\mt{^{2}\tau_{d}}$&$\mt{^{2}\al_c}$ &$\mt{^{2}\al_f}$&$\mt{^{2}t_{d}}$&$\mt{^{2}t}$&-\\ \hline
0.030 &-0.030 & -0.170 &-0.009 & 0.271 & 0.198 &- \\ \hline
$\mt{^{3}\tau}$&$\mt{^{3}\al_{2}}$&$\mt{^{3}\al_{1}}$&$\mt{^{3}t}$& - & - & -\\ \hline
-0.134&0.081&0.205&-0.260& -&-&-\\ \hline \hline
$\mt{^{4}\tau_{2}}$&$\mt{^{4}\tau_{1}}$&$\mt{^{4}\al_{f}}$&$\mt{^{4}\al}$&$\mt{^{4}\al_{c}}$&$\mt{^{4}t_{1}}$& $\mt{^{4}t_{2}}$\\ \hline
-0.090&-0.051&-0.024&-0.046&0.143&0.006&0.033\\ \hline \hline
$\mt{^{5}\tau_{2}}$&$\mt{^{5}\tau_{1}}$&$\mt{^{5}\al_{f3}}$&$\mt{^{5}\al_{f2}}$&$\mt{^{5}\al_{f1}}$&-& -\\ \hline
 0.003& 0.003&-0.054&-0.036&-0.023& - & -\\ \hline \hline
$\mt{^{5}t_{2}}$&$\mt{^{5}t_{1}}$&$\mt{^{5}\al_{c3}}$&$\mt{^{5}\al_{c2}}$&$\mt{^{5}\al_{c1}}$&-& -\\ \hline
0.051&0.048& 0.007&-0.021&-0.049& - & -\\ \hline \hline
\end{tabular}
\end{center}
\caption{Matrix elements corresponding to the hoppings near the $\al$ interface depicted in Figs. \ref{Int_hop}(a) and \ref{Int_hop}(b). These elements include first- through fifth-nearest neighbors and were extracted from Wannier analysis of the 8-$\al\bt$ system.}
\label{alpha_int}
\end{table}

\begin{figure}[b]
\includegraphics[width=0.95\columnwidth]{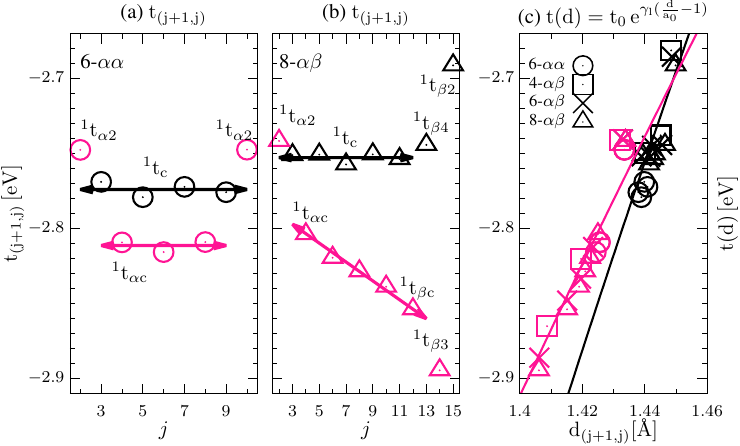}
\caption{Variation of first-nearest neighbor matrix elements $t_{(j+1,j)}$ as a function of orbital position $j$. (a) 6-$\al\al$ system and (b) 8-$\al\bt$ system. (c) $t_{(j+1,j)}$ as a function of the inter-orbital distance ($d_{(j+1,j)}$). Lines represent the fitted functional dependence [indicated in (c)], with data points corresponding to the different system sizes.}
\label{Mean_tij_03}
\end{figure}

The spatial dependence of the first-nearest neighbor matrix elements, $t_{j+1,j}$, along the graphene channel is also relevant. Figures \ref{Mean_tij_03}(a) and \ref{Mean_tij_03}(b) show $t_{j+1,j}$  vs orbital position ($j$) for the  $6$-$\al\al$ and $8$-$\al\bt$ systems, respectively, as obtained from Eq.~\eqref{AH_t1}.  A distinct separation emerges between even and odd $j$ values within the central region. Matrix elements for even $j$ exhibit a larger magnitude (more negative) than those for odd $j$;  see Fig. \ref{AH_MODEL}(c) for orbital labeling.
This behavior can be understood by the presence of internal stress within the graphene-fluorographene superlattice: fluorographene's larger lattice parameter stretches the graphene along the longitudinal axis ($\hat{x}$ direction). In fact, Fig.~\ref{AH_MODEL}(c) shows that when $j$ is even, the hopping $t_{j+1,j}$ involves orbitals separated along the $\hat{y}$ direction, while when $j$ is odd, they are separated along the $\frac{\sqrt{3}}{2}\hat{x} \pm \frac{1}{2}\hat{y}$ direction. In both cases, the distance between orbitals should be the same for a system without internal stress. However, in the superlattice of alternating graphene-fluorographene nanoribbons, the graphene channel is elongated along the longitudinal $\hat{x}$ direction, which increases the distance between orbitals that were initially in the $\frac{\sqrt{3}}{2}\hat{x} \pm \frac{1}{2}\hat{y}$ direction (odd $j$) if we compare it with the separation of orbitals along the $\hat{y}$ direction (even $j$), leading to weaker hopping terms.
To quantify this effect, Fig.~\ref{Mean_tij_03}(c) plots $t_{j+1,j}$ against the interorbital distance ($d_{j+1,j}$) obtained from relaxed DFT geometries. The exponential dependence is apparent from the figure \cite{Ribeiro_2009}.  We fit this dependence using  $t(d)=t_0\exp{(\gamma_l(\frac{d}{a_0}-1)})$, obtaining different parameters for even and odd $j$, reflecting the stress-induced anisotropy: for $j$ odd (black symbols), $t_0=-2.88$\,eV and $\gamma=-3.16$; for $j$ even (gray symbols), $t_0=-2.82$\,eV and $\gamma=-2.17$. In both cases, $a_0=1.42$\AA.
Importantly,  the value of $t_{j+1,j}$ near the edges deviates from the fitted trends of the central region. This is due to the local distortion induced by the interface. 

In summary, to model  $n$-$\al\al$ systems, we use the following system-size independent scheme for the first-nearest-neighbor hopping terms:
\begin{equation}
t_{j+1,j} = \begin{cases}
^{1}\mt{t}_{\al 1}\,, & \mbox{for } j=1,2n-1 \\
^{1}\mt{t}_{\al 2}\,, & \mbox{for } j=2,2n-2 \\
^{1}\mt{t}_{c}\,, & \mbox{for $j$ odd, and } 3\leq j\leq 2n-3 \\
^{1}\mt{t}_{\al c}\,, & \mbox{for $j$ even, and } 4\leq j\leq 2n-4
\end{cases}
\label{data_6aa}
\end{equation}
where $^1t_c=-2.774$~eV and $^1t_{\al c}=-2.811$~eV, cf.  Fig.~\ref{Mean_tij_03}(a).

The corresponding parameters for $n$-$\al\bt$ systems are
\begin{equation}
t_{(j+1,j)} = \begin{cases}
^{1}\mt{t}_{\al j}\,, & \mbox{for } j=1,2\\
^{1}\mt{t}_{\bt \nu}\,, & \mbox{for } j=2n+1-\nu\mbox{, }\nu=1,2,3,4\\
^{1}\mt{t}_{c}\,, & \mbox{for $j$ odd and } 3\leq j\leq 2n-5 \\
^{1}\mt{t}_{t}(j) \,, & \mbox{for $j$ even and } 4\leq j\leq 2n-4 \,.
\end{cases}
\label{data_8ab}
\end{equation}
In this case, the hopping term $t_{j+1,j}$ in the bulk of the channel has a constant value for $j$ odd, $^1t_c=-2.753$~eV, and a linear dependence, $^{1}\mt{t}_{t}(j)= {^{1}}\mt{t}_{\al c} + \frac{^{1}\mt{t}_{\bt c} - ^{1}\mt{t}_{\al c} }{2n-8}(j-4)$, for even $j$, with $^1t_{\bt c}=-2.854$~eV  and $^1t_{\al c}=-2.804$~eV, in accordance with Fig. \ref{Mean_tij_03}(b). 

\bibliography{ribbon_manual,ribbon_auto}

\end{document}